\documentclass[acmsmall,screen,authorversion,nonacm,review=false,timestamp=false]{acmart}

\usepackage{hyperref}

\makeatletter
\renewcommand\@formatdoi[1]{\ignorespaces}
\makeatother
\renewcommand\footnotetextcopyrightpermission[1]{} 

\acmDOI{} 
\startPage{1}
\setcopyright{none}
\renewcommand\footnotetextcopyrightpermission[1]{}
\setcopyright{none}

\makeatletter
\let\@acmArticle\@empty
\let\@authorsaddresses\@empty
\makeatother

\usepackage{booktabs}   
\usepackage{subcaption} 

\usepackage{pifont}      
\usepackage{dashbox}
\usepackage{stackengine}
\usepackage{mathpartir}
\usepackage{pftools}
\usepackage{iris}
\usepackage{wrapfig}
\usepackage[notext,nomath]{stix}

\usepackage{stmaryrd}
\usepackage[font=small]{caption}
\usepackage{siunitx}

\newcommand{\textlog}[1]{\textsf{\upshape #1}}

\DeclareFontEncoding{LS1}{}{}
\DeclareFontEncoding{LS2}{}{\noaccents@}
\DeclareFontSubstitution{LS1}{stix}{m}{n}
\DeclareFontSubstitution{LS2}{stix}{m}{n}
\DeclareSymbolFont{arrows1}{LS1}{stixsf}{m}{n}
\DeclareSymbolFont{arrows2}{LS1}{stixsf}{m}{it}
\DeclareMathSymbol{\nvrightarrowtail}{\mathrel}{arrows2}{"AD}
\DeclareMathSymbol{\nvrightarrow}{\mathrel}{arrows1}{"F6}

\newcommand{\guardsFO}{\nvrightarrow}

\newcommand{\guardsSym}{\nvrightarrowtail}

\newcommand{\preguardsSym}{\mathrel{\hat{\guardsSym}}}

\newcommand{\guards}[1][]{%
  \ifthenelse{\equal{#1}{}}{%
    \guardsSym%
  }{%
    \guardsSym_{#1}%
  }%
}%

\newcommand{\preguards}[1][]{%
  \ifthenelse{\equal{#1}{}}{%
    \preguardsSym%
  }{%
    \preguardsSym_{#1}%
  }%
}%

\NewDocumentCommand \hoareShared{O{} O{}}{
  \ifthenelse{\equal{#2}{}}{%
      {\colorShared{[#1]}}%
  }{%
      {\colorShared{[#1]_{#2}}}%
  }%
}

\newcommand{\langname}{Leaf}

\newcommand{\powset}{\mathscr{P}}

\renewcommand{\eqdef}{\triangleq}

\newcommand{\acquireExc}{\textlog{lock\_exc}}
\newcommand{\acquireShared}{\textlog{lock\_shared}}
\newcommand{\releaseExc}{\textlog{unlock\_exc}}
\newcommand{\releaseShared}{\textlog{unlock\_shared}}
\newcommand{\rwlocknew}{\textlog{rwlock\_new}}
\newcommand{\rwlockfree}{\textlog{rwlock\_free}}

\newcommand{\htnew}{\textlog{ht\_new}}
\newcommand{\htfree}{\textlog{ht\_free}}

\newcommand{\Query}{\textlog{query}}
\newcommand{\QueryIter}{\textlog{query\_iter}}
\newcommand{\Update}{\textlog{update}}
\newcommand{\UpdateIter}{\textlog{update\_iter}}

\newcommand{\uunit}{\ensuremath{\epsilon}}
\newcommand{\V}{\mathcal{V}}

\newcommand{\true}{\textsf{True}}
\newcommand{\false}{\textsf{False}}
\newcommand{\bool}{\textsf{Bool}}

\newcommand{\pointsto}{\hookrightarrow}
\newcommand{\pointstofrac}[1]{\xhookrightarrow{\smash{\raisebox{-.3ex}{\ensuremath{\scriptstyle\kern-0.25ex\textlog{frac}\kern-0.1ex}}}}_{#1}}
\newcommand{\pointstoCount}[3]{#2\xhookrightarrow{\smash{\raisebox{-.3ex}{\ensuremath{\scriptstyle\kern-0.25ex\textlog{count}\kern-0.1ex}}}}_{#1} #3}
\newcommand{\pointstoRo}[2]{#1\xhookrightarrow{\smash{\raisebox{-.3ex}{\ensuremath{\scriptstyle\kern-0.25ex\textlog{ro}\kern-0.1ex}}}} #2}

\newcommand{\stacktwo}[2]{\stackanchor{\ensuremath{#1}}{\ensuremath{#2}}}

\renewcommand{\implies}{\Rightarrow}

\renewcommand{\hoare}[3]{\ensuremath{{\{ #1 \}}~ #2 ~{\{ #3 \}}}}

\newcommand{\hash}{\ensuremath{H}}

\newcommand{\langitem}[1]{\mathtt{\color{blue}#1}}

\newcommand{\langdo}{\langitem{do}}
\newcommand{\languntil}{\langitem{until}}
\newcommand{\langrec}{\langitem{rec}}
\newcommand{\inl}{\langitem{inl}}
\newcommand{\inr}{\langitem{inr}}
\newcommand{\langfork}{\langitem{fork}}
\newcommand{\langassign}{{\color{blue}\leftarrow}}

\newcommand{\langif}{\langitem{if}}
\newcommand{\langthen}{\langitem{then}}
\newcommand{\langelse}{\langitem{else}}
\newcommand{\langmatch}{\langitem{match}}
\newcommand{\langwith}{\langitem{with}}
\newcommand{\langend}{\langitem{end}}
\newcommand{\langlet}{\langitem{let}~}
\newcommand{\langref}{\langitem{ref}}
\newcommand{\langfree}{\langitem{free}}
\newcommand{\langin}{\langitem{in}}
\newcommand{\langbang}{\langitem{!}}
\newcommand{\langabort}{\langitem{abort}}
\newcommand{\langfor}{\langitem{for}}
\newcommand{\langdone}{\langitem{done}}

\newcommand{\Some}{\textsf{Some}}
\newcommand{\None}{\textsf{None}}

\newcommand{\mapop}{\textit{m}}
\newcommand{\slotop}{\textit{slot}}

\newif\ifanon

\newcommand{\itemref}[1]{\hyperref[#1]{(\ref*{#1})}}

\newcommand{\eqnref}[1]{\hyperref[#1]{(\ref*{#1})}}
\newcommand{\exampleref}[1]{\hyperref[#1]{Example~\ref*{#1}}}

\makeatletter
\def\sectionautorefnamesymbol{\S\@gobble}

\makeatother

\newcommand{\mle}{\preceq}
\newcommand{\mge}{\succeq}

\newcommand{\Fields}{\textsf{Fields}}
\newcommand{\ExcPending}{\textsf{ExcPending}}
\newcommand{\SharedPending}{\textsf{ShPending}}
\newcommand{\ExcGuard}{\textsf{Exc}}
\newcommand{\SharedGuard}{\textsf{Sh}}

\newcommand{\elemFields}{\textsf{fields}}
\newcommand{\elemExcPending}{\textsf{excPending}}
\newcommand{\elemSharedPending}{\textsf{shPending}}
\newcommand{\elemExcGuard}{\textsf{exc}}
\newcommand{\elemSharedGuard}{\textsf{sh}}

\newcommand{\RwFamily}{\textsf{RwFamily}}

\definecolor{colshared}{HTML}{8C368C} 
\newcommand{\colorShared}{\color{colshared}}

\newcommand{\CAS}{\langitem{CAS}}
\newcommand{\FetchAdd}{\langitem{FetchAdd}}

\newcommand{\inlinehoaresmall}[1]{{\{#1\}}}
\newcommand{\inlinehoare}[1]{{\left\{#1\right\}}}

\newcommand{\langindent}{\phantom{aa}}

\newcommand{\MediumAsterisk}{\mathop{\scalebox{2.2}{\raisebox{-0.2ex}{$\ast$}}}}%

\newcommand{\rulelabel}[1]{\tag{\rulenamestyle{#1}}\label{#1}}
\newcommand{\rulelabelx}[1]{\tag{\rulenamestyle{#1}}\label{X#1}}

\definecolor{darkgreen}{RGB}{25,99,45}

\newcommand{\inde}{\langindent\langindent}

\newcommand{\IsRwLock}{\textlog{IsRwLock}}
\newcommand{\IsHT}{\textlog{IsHT}}

\newcommand{\rw}{\textit{rw}}
\newcommand{\htable}{\textit{ht}}
\newcommand{\locks}{\textit{locks}}
\newcommand{\slots}{\textit{slots}}

\newcommand{\Set}{\textit{Set}}
\newcommand{\Value}{\textit{Value}}
\newcommand{\Key}{\textit{Key}}
\newcommand{\Name}{\textit{Name}}

\newcommand{\entails}{\vdash}

\newcommand{\bothentails}{\dashv\vdash}

\newcommand{\Inv}{\mathcal{C}}
\newcommand{\Sfunc}{\mathcal{S}}

\newcommand{\dispdothoriz}[2][3mu]{\dot{#2\mkern#1}\mkern-#1}
\newcommand{\dispdothorizLeft}[2][3mu]{\dot{#2\mkern-#1}\mkern#1}

\newcommand{\rightdot}[1]{\dispdothoriz[14mu]{#1}}
\newcommand{\leftdot}[1]{\dispdothorizLeft[14mu]{#1}}

\newcommand{\exchange}{\mathbin{\leftdot{\rightsquigarrow}\hspace{-1.1em}\rightdot{\phantom{\rightsquigarrow}}}}
\newcommand{\deposits}{\mathbin{\leftdot{\rightsquigarrow}}}
\newcommand{\withdraws}{\mathbin{\rightdot{\rightsquigarrow}}}

\newcommand{\transitions}{\rightsquigarrow}

\DeclareMathOperator{\maps}{sto}

\newcommand{\protOwn}[2]{\langle{#1}\rangle^{#2}}

\DeclareMathOperator{\RespectsComposition}{RespectsComposition}

\newcommand{\AllocatedInvs}{\mathsf{AllocatedInvs}}
\newcommand{\OwnInvs}{\mathsf{OwnInvs}}

\newcommand{\init}{\mathsf{prot}}
\newcommand{\uninit}{\uunit}
\newcommand{\Prot}{\textsc{Prot}}

\newcommand{\exc}{\textit{exc}}
\newcommand{\rc}{\textit{rc}}

\newcommand{\gammaht}{\gamma_\text{ht}}

\renewcommand{\Auth}{\textsc{Auth}}
\newcommand{\Ex}{\textsc{Excl}}
\newcommand{\AgN}{\textsc{AgN}}

\newcommand{\UnitType}{\mathbf{1}}

\newcommand{\exConstructor}{\textsf{ex}}
\newcommand{\agnConstructor}{\textsf{agn}}

\newcommand{\ePending}{\textit{ep}}
\newcommand{\sPending}{\textit{sp}}

\newcommand{\agncount}{\textit{count}}

\newcommand{\NamespaceForever}{{\mathcal{F}\mkern-2mu\textit{orever}}}

\newcommand{\NamespaceFrac}{{\mathcal{F}\mkern-2mu\textit{rac}}}
\newcommand{\NamespaceCount}{{\mathcal{C}\mkern-2mu\textit{ount}}}

\newcommand{\pointprop}[1]{\textlog{point}(#1)}

\newcommand{\HeapInterp}{\mathcal{H}}

\DeclareMathOperator{\KeysDistinct}{KeysDistinct}
\DeclareMathOperator{\MapConsistent}{MapConsistent}
\DeclareMathOperator{\SlotsConsistent}{SlotsConsistent}
\DeclareMathOperator{\Contiguous}{Contiguous}

\newcommand{\appref}[1]{\hyperref[#1]{Appendix~\ref*{#1}}}

\begin{document}
\anontrue
\begin{anonsuppress}
\anonfalse
\end{anonsuppress}

\title[Leaf: Modularity for Temporary Sharing in Separation Logic]{Leaf: Modularity for Temporary Sharing in Separation Logic (Extended Version)}


\author{Travis Hance}\affiliation{\institution{Carnegie Mellon University}\country{USA}}\email{thance@andrew.cmu.edu}
\author{Jon Howell}\affiliation{\institution{VMware Research}\country{USA}}\email{howell@vmware.com}
\author{Oded Padon}\affiliation{\institution{VMware Research}\country{USA}}\email{oded.padon@gmail.com}
\author{Bryan Parno}\affiliation{\institution{Carnegie Mellon University}\country{USA}}\email{parno@cmu.edu}

\makeatletter
\let\@authorsaddresses\@empty
\makeatother

\makeatletter
\def\@listi{\leftmargin\leftmargini
    \parsep 1\p@ \@plus0\p@ \@minus\p@
    \topsep 2\p@   \@plus0\p@ \@minus\p@
    \itemsep1\p@ \@plus0\p@ \@minus\p@}
\let\@listI\@listi\@listi
\makeatother

\begin{abstract}
In concurrent verification, separation logic provides a strong story for handling both 
resources that are owned exclusively and resources that are
shared persistently (i.e., forever).
However, the situation is more complicated for temporarily shared state, where state might be shared
and then later reclaimed as exclusive.
We believe that a framework for temporarily-shared state should meet two key goals
not adequately met by existing techniques.
One, it should allow and encourage users to verify new sharing strategies.
Two, it should provide an abstraction where users manipulate shared state in a way
  agnostic to the means with which it is shared.

We present {\langname}, a library in the Iris separation logic
which accomplishes both of these goals
by introducing a novel operator, which we call \emph{guarding}, that allows one
proposition to represent a shared version of another.
We demonstrate that {\langname} meets these two goals through a modular case study:
we verify a reader-writer lock that supports shared state, and a hash table built on top of it
that uses shared state.

\end{abstract}



\maketitle

\section{Introduction}\label{sec-intro}

Multi-threaded concurrent programs are difficult to get right.
One challenging pattern in such programs is \emph{read-sharing},
i.e., allowing multiple threads to simultaneously read mutable shared state 
as long as no other thread is actively writing.
This common optimization reduces thread contention and is often considered
critical for scaling concurrent performance.
While the general idea is commonly deployed,
the concrete instantiations vary wildly.
For example, even the implementation of a conceptually simple reader-writer lock
grows quite complicated as various kinds of scaling issues are considered~\cite{scalable-rwlocks-for-parallel-systems,lock-unlock-is-that-all,BRAVO,shirako-rwlock-model-checking,scalable-numa-aware,scalable-read-mostly,numa-aware-rwlocks}.
As a concrete instance, one possible optimization uses multiple reference counters, each on its own cache line, to reduce thread contention for readers~\cite{numa-aware-rwlocks}.
And yet the challenge does not stop at reader-writer locks; for instance, the node-replication algorithm of \citet{nr-asplos17} might allow simultaneous read-access to particular entries
of a ring buffer, and this protocol does not resemble a lock in the slightest.
Given all of this complexity,
we would naturally like to verify, in a modular fashion,
both the read-sharing implementations and the programs that use them.

One effective tool for reasoning about concurrent programs is \emph{concurrent separation logic (CSL)}~\cite{separationlogic,resources-concurrency-local-reasoning}
which lets us reason naturally about exclusive ownership of memory and, more generally, arbitrary resources.
This is a great fit for (mutually exclusive) locks: it is easy to write a specification of a lock that allows the client to obtain ownership of a resource
(e.g., permission to access some part of memory, or a more complicated invariant) which is returned upon releasing the lock, allowing the resource to be transferred
between threads.

But how do we handle simultaneous read-sharing?
We want some way of reasoning about this \emph{simultaneous, shared} access---e.g., we want to talk about memory access where we can only read but not write,
or invariants which must be preserved until the shared access is released.
CSL's exclusive ownership does not work here, since the desired type of ownership is not exclusive.
Some more recent CSLs also have a concept called \emph{persistent knowledge}~\cite{OnModelsOfHigherOrderSL}, which allows indefinite sharing, but this on its own does not suffice either:
whatever sharing mechanism we use needs a way to \emph{reclaim} exclusive access once all shared access has been revoked.

Two of the earliest proposed methods to represent shared state while allowing reclamation are
\emph{fractional permissions}~\cite{boyland-fractional} and \emph{counting permissions}~\cite{permission-accounting-in-separation-logic}.
These representations have the benefit of being easy to construct and prove sound, but they do not form complete proof strategies for complex sharing protocols like the above,
which are likely to have considerably more state that evolves in complex ways and cannot be expressed through fractional permissions or reference counters alone.
Furthermore, the desire for modularity means there is pressure for specifications to converge on one particular representation so that they can interoperate; this limits flexibility
since different representations might be more suitable for different applications.
How can we support a wide gamut of sharing techniques---so the user can choose the right tool for the job, even building their own representation if necessary---while also achieving modularity?

Our approach is to take the core idea of fractional and counting permissions, generalize it, and a provide a uniform way to reason about it.
That core idea, we argue, is that both involve propositions which are able to ``stand in'' for some kind of shared resource,
but which are suitable for manipulation within the substructural separation logic. This enables them to handle the temporality of the sharing.
This motivates us to extract the essence of this ``standing-in'' relationship and brings us to our contribution, the {\langname} logic.

{\langname} introduces a novel operator $\guards$, which we call \emph{guarding}, to represent the relationship between an exclusively owned proposition and the shared proposition it represents.
{\langname} handles both sides of this abstraction. First, we show how the user can \emph{deduce} nontrivial $\guards$ relationships
by constructing arbitrary sharing protocols. Such protocols include ones based on the above-mentioned patterns, as well as
custom protocols that are tailored to particular implementations. Our approach is inspired by \emph{custom ghost state}~\cite{CommSTSFineGrained,iris-2015,views-compositional,superficially-substructural-types,subjective-auxiliary-state,ConcAbsPred},
a class of flexible separation logic techniques.
We call the protocols of our new formulation \emph{storage protocols}.
Second, we show how to \emph{make use of} $\guards$ relationships,
through general rules that are agnostic to the underlying sharing mechanism,
thus enabling modular specifications.
This approach is symbiotic between ghost state and read-sharing:
by applying custom-ghost-state techniques, Leaf
supplies a general form for read-sharing mechanisms; 
meanwhile, some ghost-state constructions become simpler because they can rely
on Leaf's read-sharing, without needing to include their own bespoke sharing mechanisms.

Storage protocols can support complex algorithms found in real-world concurrent software systems.
As we discuss in \autoref{sec:advanced}, {\langname}'s storage protocols have already been used in the 
IronSync framework~\cite{ironsync-tr}, which is enabled by Leaf's systematic approach to
read-shared custom ghost state.
IronSync targets production-scale, high-performance concurrent systems, e.g.,
a multi-threaded page cache (reaching 3M ops / second)
or the node-replication algorithm mentioned above (3M ops / second across 192 threads).
These results confirm that high-performance applications contain sophisticated,
domain-specific read-sharing patterns;
demonstrate that Leaf's storage protocols can handle them; 
and provide evidence that system developers find Leaf's perspective on temporarily read-shared 
resources useful.
In this paper, though, we will mostly focus on the technical formalism of this perspective, so we use a smaller, self-contained example that is simple enough to explain in full, but still complex enough to show the utility of Leaf.

Specifically, we make the following contributions:
\begin{itemize}
  \item We present {\langname}, which has
      built-in deduction rules for temporarily shared state and a mechanism for user-defined sharing protocols based on ghost state, which we call \emph{storage protocols}.
  \item We show how storage protocols capture existing patterns, e.g., fractional and counting permissions.
  \item We illustrate {\langname}'s modular specifications through a case study of a reader-writer lock and a hash table, demonstrating several different facets of sharing:
    \begin{itemize}
      \item The hash table itself is shared between threads.
      \item The hash table is composed of many reader-writer locks, which inherit that sharedness.
      \item The memory cells in the reader-writer lock further inherit that sharedness, and they are accessed atomically by multiple threads.
      \item The reader-writer locks allow temporary, shared read-only access to the hash table's memory slots. Furthermore, this is done via a clean, modular specification of the reader-writer lock.
    \end{itemize}
  \item We prove the soundness of {\langname} and provide it as a library in the Iris framework~\cite{iris-from-the-ground-up} in Coq.\footnote{\href{https://github.com/secure-foundations/leaf}{https://github.com/secure-foundations/leaf}}
      We mechanize our case studies in Coq. These proofs are available as open source
      and in our supplementary materials.
\end{itemize}

\section{Overview}

{\langname} is constructed as a library in the \emph{Iris separation logic}~\cite{iris-from-the-ground-up}.
We generally use Iris notation, where applicable, and all the standard Iris proof rules apply.
We assume familiarity with separation logic basics (the connectives $*$, $\wand$, and so on), but we will review key Iris features as they come up.

\subsection{Leaf Introduction: Resource Guarding}

The primary question we need to unravel is how to talk generally about ``a shared $P$'' for any proposition $P$.
Here, the proposition $P$ might be something simple, like the permission to access a certain memory location $\ell$ and read a specific value $v$,
denoted $\ell \pointsto v$, or it might be a more complex invariant.
In order to make shared state reclaimable, we connect the ``shared $P$'' to some exclusively owned (not persistent) proposition.

We do this via a relationship $G \guards P$, pronounced \emph{$G$ guards $P$}.
$G \guards P$ is itself a proposition; informally, it means that $G$ can be used as a ``shared $P$.''
Hence, if some program proof needs to operate over a ``shared $P$,'' it can instead take $G$ as an exclusively owned precondition,
and use the relationship $G \guards P$ when it needs to use $P$. Later, $G$ might be consumed (disallowing further shared access to $P$),
and eventually the exclusive ownership of $P$ might be reclaimed.

In general, then, when we want to write a proof that operates over some read-only $P$
in a way that abstracts over the way $P$ is shared,
we can write the proof to take ownership of some arbitrary Iris proposition $G: \iProp$ where $G \guards P$.
To codify this pattern, we use a shorthand,
$\hoareShared[ X ]~\{ P \}~e~\{ Q \}$, to mean
$\forall G: \iProp.~ \{ P * G * (G \guards {\colorShared X}) \}~e~\{ Q * G \}$.
This can be read as ``if command $e$ executed, with $P$ owned at the beginning,
and with $\colorShared X$ shared, then $Q$ is owned at the end.''
We {\colorShared indicate shared resources in purple}, though this is only a visual aid, and it has no syntactic meaning.

For example, a program logic might allow writing to a memory location given exclusive ownership of $\ell \pointsto v$,
but allow reading from it given \emph{shared} ownership of $\ell \pointsto v$.
{\langname} specifies this as:
{\small\begin{mathpar}
    \axiomhref{Heap-Write}{Overview-Heap-Write}{
      \{ \ell \pointsto v \} ~ \ell ~\langassign~ v' ~ \{ \ell \pointsto v' \}
    }

    \axiomhref{Heap-Read-Shared}{Overview-Heap-Read-Shared}{
      \hoareShared[\ell \pointsto v]~
      \{  \} ~ \langbang\ell ~ \{ r.~ v = r \}
    }
\vspace{-2mm}
\end{mathpar}}%
Here, $\ell ~\langassign~ v'$ is the command to write to the reference $\ell$, while $\langbang \ell$ reads it.
A bound variable in a postcondition, e.g.~$r$ here, represents the command's return value, so \ruleref{Overview-Heap-Read-Shared} 
says, if we have a shared $\colorShared \ell \pointsto v$ and read from $\ell$, then we obtain a value equal to $v$.

\subsection{Example: A Reader-Writer Lock Specification}

The reader-writer lock spec in \autoref{fig:rwlock:spec} illustrates several facets
of our guarding system.
The API of this lock has six functions: $\rwlocknew$ and $\rwlockfree$ are the constructor
and destructor, respectively; $\acquireExc$ and $\releaseExc$
are intended to allow exclusive, write access to some underlying resource;
$\acquireShared$ and $\releaseShared$ are intended to allow shared, read-only access.
Exactly what this ``resource'' is may be determined by the client.

Holding the spec together is the proposition $\IsRwLock(\rw, \gamma, F)$, which roughly says
that the value $\rw$ is a reader-writer lock with a unique identifier $\gamma$.
$F$ is used to specify the resource being protected---we will return to this in a moment.
Note that when a new reader-writer lock is constructed (via $\rwlocknew$)
the client obtains exclusive ownership over $\IsRwLock(\rw, \gamma, F)$;
on the other hand, the operations that are meant to run concurrently all take
$\colorShared \IsRwLock(\rw, \gamma, F)$ as \emph{shared}.
The destructor, $\rwlockfree$, again requires non-shared ownership, as naturally it should not be able to run concurrently with other operations.

\begin{figure}
\begin{center}\small
    \textbf{RwLock Specification}

    \textbf{Propositions:}\quad
      $\IsRwLock(\rw, \gamma, F)$\quad
      $\ExcGuard(\gamma)$\quad
      $\SharedGuard(\gamma, x)$\quad

    \text{(where $\rw: \Value,~~ \gamma: \Name,~~ X: \Set,~~ x: X,~~ F: X \to \iProp$)}

  \newcommand{\hackhback}{\hspace{-0.15in}}
  \vspace{-4mm}
  \small\begin{align*}
    \forall F,x.~ &&
    \{ F(x) \}~&\rwlocknew()&&~\hackhback\{ \rw .~ \exists\gamma, \IsRwLock(\rw, \gamma, F) \} &
    \\
    \forall \rw, \gamma, F.~ &&
    \{ \IsRwLock(\rw, \gamma, F) \}~&\rwlockfree(\rw)&&~\hackhback\{ \} &
    \\
    \forall \rw,\gamma,F .~ &
    \hoareShared[ \IsRwLock(\rw, \gamma, F) ]~&
    \{ \}~&\acquireExc(\rw)&&~\hackhback\{ \ExcGuard(\gamma) * \exists x .~ F(x) \} &
    \\
    \forall \rw,\gamma,F,x .~ &
    \hoareShared[ \IsRwLock(\rw, \gamma, F) ]~&
    \{ \ExcGuard(\gamma) * F(x) \}~&\releaseExc(\rw)&&~\hackhback\{ \} &
    \\
    \forall \rw,\gamma,F .~ &
    \hoareShared[ \IsRwLock(\rw, \gamma, F) ]~&
    \{ \}~&\acquireShared(\rw)&&~\hackhback\{ \exists x.~ \SharedGuard(\gamma, x) \} &
    \\
    \forall \rw,\gamma,F,x .~ &
    \hoareShared[ \IsRwLock(\rw, \gamma, F) ]~&
    \{ \SharedGuard(\gamma, x) \}~&\releaseShared(\rw)&&~\hackhback\{ \} &
    \\
    \forall \rw,\gamma,F,x.~ &
    \IsRwLock(\rw, \gamma, F) \entails (\SharedGuard(\gamma, x) \guards {\colorShared F(x)}) \hspace{-1in} &&&&&
  \end{align*}
\end{center}
  \vspace{-4mm}
\caption{ \label{fig:rwlock:spec} \small
  Example specification for a reader-writer lock using {\langname} notation.
  In \autoref{sec:rwlock}, we show how to prove this specification for a particular
  implementation.
}
  \vspace{-4mm}
\end{figure}

Now, the client needs to specify what sort of resource they want to protect.
For example, the client might want to protect access to some location in memory, say $\ell$,
so they would use the lock to protect resources of the form $\ell \pointsto v$.
To allow the client to choose the kind of resource they want to protect,
our specification lets the client, upon construction of the lock, provide
a \emph{proposition family} $F : X \to \iProp$ parameterized over some set $X$.
In the above example, we might have $F = \lambda x : \Value .~ \ell \pointsto x$ for some fixed $\ell$
determined at the time of the $\rwlocknew()$ call.

In the specification, observe that we then use $F(x)$, for some $x$,
to represent the resource when it is obtained from the lock by $\acquireExc$.
Upon calling $\releaseExc$,
the client then has to return some $F(x')$, where $x'$ might be different than $x$.
This makes sense, because $\acquireExc$ is supposed to be a write-lock, so the client
should be able to manipulate the given resource at will, provided it restores the lock's
invariants.

Acquiring the shared lock is more interesting, since we have to acquire some
$F(x)$ resource in a \emph{shared} way. This is where the $\guards$ operator
comes in: rather than receiving $F(x)$ directly, the client obtains a special
resource $\SharedGuard(\gamma, x)$ (for some $x$), for which we have
$\SharedGuard(\gamma, x) \guards \colorShared F(x)$.
Thus, the client has shared access to $\colorShared F(x)$ as long as it has the $\SharedGuard$, which
must be relinquished upon release of the lock.
We view $\SharedGuard$ as a separation logic analogue of a \emph{lock guard}, an object in some locking APIs~\cite{rust-lock-guard,cpp-lock-guard} which exists for the duration of a held lock.
Indeed, this inspires the \emph{guarding} name. 

Notice the choice of parameter set $X$ and proposition family $F$ determines exactly what it means to be ``read-only,''
because it is the value $x : X$ which is fixed until the client releases the shared lock.
For example, we might set $X = \mathbb{Z}$ and $F = \lambda x : \mathbb{Z} .~ \ell \pointsto x$. Then the client can take a shared lock and obtain shared $\colorShared \ell \pointsto x$ for some fixed integer $x$, which cannot change
until the lock is released.
On the other hand, they might set, say, $X = \mathbb{Z}_2$ and $F = \lambda x: \mathbb{Z}_2 .~ \exists n.~ (\ell \pointsto n) * (n = x \mod 2)$.
In this case, upon taking the shared lock, they receive $\colorShared \ell \pointsto v$, but now only the \emph{parity} of $v$ is fixed. In fact, in this situation, the user would be able
to update $v$ to another value of the same parity (provided they do so in an atomic operation).

The RwLock spec raises two questions:
How can the client do interesting things when they have $\SharedGuard(\gamma, x)$, i.e., a ``shared $\colorShared F(x)$'', rather than exclusive ownership of $F(x)$?
Secondly, how can we verify a realistic lock implementation against this spec,
which requires the deduction of a nontrivial guard relationship
$\SharedGuard(\gamma, x) \guards \colorShared F(x)$?
Let us tackle these in turn.

\subsection{Utilizing Shared State}

How does the user actually benefit from shared state, i.e., state (like $\colorShared F(x)$) under a guard operator,
as in $(\SharedGuard(\gamma, x) \guards {\colorShared F(x)})$
from the previous example?

In general, if $G \guards P$, {\langname} aims to let $G$ be usable in any operation
that could have used $P$, provided that $P$ is not modified.
Such an operation might be given by the Iris operator called the \emph{view shift}, as in $P * A \vs P * B$.
In general, the view shift $(\vs)$ effectively says we can give up the resources on the left side to obtain the resources on the right.
In the example, though, with $P$ on both the left and right sides, $P$ is \emph{not} consumed, although it \emph{is} needed to perform the operation.
In this case, we could use $G$ in place of $P$; i.e., we would have $G * A \vs G * B$.

Frequently, in order to perform such updates,
we need to first compose multiple pieces of shared state together; for example, suppose we employ a fine-grained locking scheme,
where a thread might hold multiple pieces of state from different locks in shared mode,
which \emph{all together} are needed to perform a certain update or deduction.
Ordinarily, we would compose the corresponding propositions with separating conjunction ($*$),
but here, the pieces, being shared, might come from the same source and not actually be separated.
To get around this, we use overlapping conjunction ($\land$) rather than $*$ when dealing with shared state.
It turns out that constructing sound deduction rules to use $\land$ is subtle; the rule we give in \autoref{sec:logic:main} requires a specific technical condition.
We will show how all this works together through our fine-grained hash table example (\autoref{sec:ht}).

\subsection{Deducing Guard Relationships}

Towards verifying an implementation of a reader-writer lock, the most salient technical question is
how we can construct nontrivial propositions like $\SharedGuard(\gamma, x)$
and prove guard relationships on them.
To tackle this question, it helps to first look at simpler examples of nontrivial guard relationships.
As such, let us take a look at
\emph{fractional permissions}~\cite{boyland-fractional} and \emph{counting permissions}~\cite{permission-accounting-in-separation-logic}, two of the oldest known
methods used to account for reclaimable read-shared permissions for memory (and other resources).

\begin{figure}
  \begin{subfigure}[t]{0.5\textwidth}
    \begin{center}\small
        \textbf{Fractional Permissions}
        \vspace{0.2em}

        \textbf{Propositions:}\quad
            $\ell \pointstofrac{q} v$

        (where $\ell : \Loc,~~ v: \Value,~~ q : \mathbb{Q}$, and $q > 0$)
        \begin{mathpar}
        (\ell \pointsto v) \vsE[\NamespaceFrac] (\ell \pointstofrac{1} v)

        (\ell \pointstofrac{q} v) \guards[\NamespaceFrac] (\ell \pointsto v)

        (\ell \pointstofrac{q_1 + q_2} v) \bothentails
            (\ell \pointstofrac{q_1} v) * (\ell \pointstofrac{q_2} v)
        \end{mathpar}
    \end{center}
  \end{subfigure}
  ~
  \begin{subfigure}[t]{0.5\textwidth}
    \begin{center}\small
        \textbf{Counting Permissions}
        \vspace{0.2em}

        \textbf{Propositions:}\quad
            $\pointstoCount{n}{\ell}{v}$
            \quad
            $\pointstoRo{\ell}{v}$

        (where $\ell : \Loc,~~ v: \Value,~~ n : \mathbb{N}$)
        \begin{mathpar}
        (\ell \pointsto v) \vsE[\NamespaceCount] (\pointstoCount{0}{\ell}{v})

        (\pointstoRo{\ell}{v}) \guards[\NamespaceCount] (\ell \pointsto v)

        (\pointstoCount{n}{\ell}{v}) \bothentails
            (\pointstoCount{n + 1}{\ell}{v}) * (\pointstoRo{\ell}{v})
        \end{mathpar}
    \end{center}
  \end{subfigure}
  \vspace{-3mm}
    \caption{ \small
        \label{fig:frac:count:logic}
        Fractional and counting permissions expressed by the $\guards$ operator.
        The $\vsE$ means we can perform an \emph{update} to
        exchange exclusive ownership of one side for the other,
        while $\bothentails$ means both sides are equivalent.
        $\NamespaceFrac$ and $\NamespaceCount$ are arbitrary
        \emph{namespaces}~(\autoref{sec:protocol}).
        In {\langname}, these laws are derived from
        storage protocols (\autoref{sec:protocol}).
    }
  \vspace{-3mm}
\end{figure}

\subsubsection{Fractional and Counting Examples}
In the fractional paradigm, the points-to proposition is labeled with a rational number $q : Q$.
These propositions combine additively:
$(\ell \pointstofrac{q} v) * (\ell \pointstofrac{q'} v) ~~\bothentails~~ (\ell \pointstofrac{q + q'} v)$,
where the $\bothentails$ is bidirectional entailment, i.e., the two sides are equivalent.
Write permission is given by $\ell \pointstofrac{1} v$ and read permission is given by
any $\ell \pointstofrac{q} v$ where $q > 0$.
The idea is that the $\ell \pointstofrac{1} v$ can be split into multiple fractional pieces,
which can be handed out and used in a read-only fashion, and then put back together to obtain
write access, allowing the user to change $v$.
Intuitively, the reason this works is that one cannot reclaim write access without gathering
\emph{all} the read-only pieces, since all of them are needed to sum back to $1$.
Thus anyone holding onto a read-only piece cannot have the value changed out from under
them by another thread.

Counting permissions, on the other hand, does not allow arbitrary splitting, but instead uses a centralized counter, $\pointstoCount{n}{\ell}{v}$ ($n : \mathbb{N}$)
to keep track of the number of extant read-only permissions, denoted $\pointstoRo{\ell}{v}$.
The user can increment the counter to obtain another read-only permission, or perform the inverse:
$(\pointstoCount{n}{\ell}{v})
  ~~\bothentails~~
  (\pointstoCount{n+1}{\ell}{v}) * (\pointstoRo{\ell}{v})$.
Meanwhile, $\pointstoCount{0}{\ell}{v}$ gives write permission; i.e., we can write as long
as there are zero read permissions in existence.

In {\langname}, we can express both these patterns using $\guards$, as shown in
\autoref{fig:frac:count:logic}.
The idea of the ``write permission'' is expressed by saying that $\ell \pointstofrac{1} v$
can be exchanged for $\ell \pointsto v$, and vice versa; the ``read permission''
is expressed by the guards relationship:
$(\ell \pointstofrac{q} v) \guards[\NamespaceFrac] (\ell \pointsto v)$.
(We explain the $\NamespaceFrac$ label later.)
The same approach works for the counting permissions.

Note that with this setup, we do not need to prove the heap read and write rules for
the fractional and counting permissions individually. Rather, we simply apply the more general
\ruleref{Overview-Heap-Write} and \ruleref{Overview-Heap-Read-Shared} rules from earlier
along with any guard relationship, such as one from \autoref{fig:frac:count:logic}.

\subsubsection{Nontrivial Guarding with Storage Protocols}

Now, we return to our question from earlier: how can we, in general,
soundly construct propositions like $(\ell \pointstofrac{q} v)$, $(\pointstoRo{\ell}{v})$,
or $\SharedGuard(\gamma, x)$?
What are the primitive deduction rules for $\guards$, and in particular, what are the rules
that allow us to prove nontrivial $\guards$ relations on those propositions?

All of these can be constructed by via a {\langname} formalism called a \emph{storage protocol},
so named because they allow the user to ``store'' propositions (e.g., by 
$(\ell \pointsto v) \vs[\NamespaceFrac] (\ell \pointstofrac{1} v)$)
and then access them in a shared manner.
The core idea is based off of \emph{custom ghost state}, a concept wherein a user
may define their own resources and derive update relations $(\vs)$.
Storage protocols extend the concept to also allow the derivation of guard relations $(\guards)$.

The propositions constructed by the protocol are able to guard arbitrary propositions that have no intrinsic notion of being shareable.
For example, in the fractional example, $\ell \pointsto v$ has no notion of being shareable;
rather, \emph{given} $\ell \pointsto v$, without knowing anything about its definition, {\langname} allows us to
construct the $\ell \pointstofrac{q} v$ proposition with a particular guard relationship to $\ell \pointsto v$.
This is an instance of the same feature that lets us have $\SharedGuard(\gamma, x) \guards F(x)$ parameterized by an arbitrary proposition family $F$.

\subsection{Outline}

Throughout the paper, we explore these examples in more detail.
We first formally introduce $\guards$ and its elementary deduction rules, and sketch how
we can derive rules like \ruleref{Overview-Heap-Read-Shared} within {\langname}.
We then 
introduce our new formulation of custom ghost state, allowing the deduction of
nontrivial $\guards$ propositions, such as those in \autoref{fig:frac:count:logic}.
We show how to verify a reader-writer lock, proving the specification (\autoref{fig:rwlock:spec}) holds for
a particular implementation. To illustrate {\langname}'s modular specifications, we 
then build another application on top of the reader-writer lock.
Finally, we discuss our construction of {\langname} within Iris and the definition of $\guards$, proving our laws sound.

\section{The {\langname} Logic} \label{sec:logic}

We begin our presentation by reviewing the concept of custom ghost state.
The formulation we present first is (largely) standard, yet still significant within {\langname}.
Then we will dive into {\langname}'s $\guards$ operator
and our new extension of custom ghost state.

\subsection{Custom Ghost State (Background)} \label{sec:bg:pcm}

\emph{Custom ghost state} in Iris is a mechanism through which the user can soundly construct their own resource with custom update rules.
We present a simplified version of it here, based primarily on the ``Iris 1.0'' formulation~\cite{iris-2015}.

The construction is parameterized by a \emph{partial commutative monoid} (PCM) whose elements form the basis of the resource.
Formally, a PCM is a set $M$ (the \emph{carrier} set) with a composition operator $\cdot : M \times M \to M$ which is associative and commutative, and with a unit element $\uunit$.
We let $a \mle b \eqdef (\exists c .~ a\cdot c = b)$.
The partiality is represented by a
\emph{validity predicate} $\V : M \to \bool$, where $\V(\uunit)$ and $\forall a,b.~ a \mle b \land \V(b) \implies \V(a)$.
We also define a derived relation $\transitions$ called the \emph{frame-preserving update},
{\small\begin{align*}
  a \transitions b \eqdef \forall c .~ \V(a \cdot c) \implies \V(b \cdot c).
\end{align*}}Essentially, $a$ can transition to $b$ if for any valid way of ``completing'' state,
the state would remain valid after the transition.

For any such $M$, Iris shows the rules in \autoref{fig-pcm-basic-rules} are sound
for a proposition written $\ownGhost{\gamma}{a}$ for $a : M$.
The $\gamma: \Name$ is a \emph{ghost name} (sometimes called \emph{ghost location})
from an arbitrary, infinite set of available names.
These rules show, for instance, that the compositional structure $(\cdot)$ of the monoid determines the compositional structure within the logic,
i.e., $\ownGhost{\gamma}{a \cdot b}$ is equivalent to $\ownGhost{\gamma}{a} * \ownGhost{\gamma}{b}$.

Likewise, an update $a \transitions b$ means that we can exchange $\ownGhost{\gamma}{a}$ for $\ownGhost{\gamma}{b}$ as resources within the logic:
$\ownGhost{\gamma}{a} \vs \ownGhost{\gamma}{b}$ as given by \ruleref{PCM-Update}.
The operator $\vs$ is called \emph{view shift}, and it essentially means we can give up the resource on the left-hand to obtain the resource on the right.
\ruleref{VS-Hoare} says we can perform such updates at any program point during a proof.
Note that the view shifts can also be annotated with a \emph{mask}, denoted $\vs[\mask]$;
we discuss this further in the next section.

\begin{example}
  An archetypal PCM is the \emph{exclusive monoid}, $\Ex(X)$, for a given set $X$.
  The elements of $\Ex(X)$ are made out of the following symbols:
  {\small\begin{align*}
      \uunit ~~|~~ \exConstructor(x) ~~|~~ \mundef
      \quad\quad
      \text{with $\forall x,y.~ \exConstructor(x) \cdot \exConstructor(y) = \mundef$ and $\forall a$, $a \cdot \uunit = a$ and $a \cdot \mundef = \mundef$}
  \end{align*}}%
      %
      Here, $\uunit$ is the unit element, representing ownership of nothing,
      the value $\exConstructor(x)$ represents exclusive ownership of a state $x$,
      and $\mundef$ represents the impossible ``conflict'' state of multiple ownership claims.
      The elements $\uunit$ and $\exConstructor(x)$ are all considered ``valid,'' while $\mundef$ is ``invalid,''
      i.e., $\V(\mundef) = \false$.
      One can show that for any $x, y : X$,
      $\exConstructor(x) \transitions \exConstructor(y)$, which implies the view shift,
      $\ownGhost{\gamma}{\exConstructor(x)} \vs \ownGhost{\gamma}{\exConstructor(y)}$ by \ruleref{PCM-Update}.
      That is, given ownership of the state,
      one can freely update it.
\end{example}

\begin{figure}
  \begin{subfigure}[t]{.62\textwidth}
  \begin{center}\small
      \textbf{Rules for PCM ghost state}

      Instantiated for a given PCM $(M, \cdot, \V)$

      \textbf{Propositions:}\quad
          $\ownGhost{\gamma}{a}$
          \quad
      \text{(where
      $a : M,~~
      \gamma : \Name$)}

    \begin{mathpar}
      \axiomhref{PCM-Unit}{PCM-Unit}{
      \true \entails \ownGhost{\gamma}{\uunit}
      }

      \axiomhref{PCM-Valid}{PCM-Valid}{
      \ownGhost{\gamma}{a} \entails \V(a)
      }

      \axiomhref{PCM-Sep}{PCM-Sep}{
      \ownGhost{\gamma}{a\cdot b}\bothentails\ownGhost{\gamma}{a} * \ownGhost{\gamma}{b}
      }

      \inferhref{PCM-Alloc}{PCM-Alloc}{
        \V(a)
      }{
      \true \vs \exists\gamma.~ \ownGhost{\gamma}{a}
      }

      \inferhref{PCM-Update}{PCM-Update}{
        a \transitions b
      }{
        \ownGhost{\gamma}{a} \vs \ownGhost{\gamma}{b}
      }

      \inferhref{PCM-And}{PCM-And}{
          \forall t.~ ((x \mle t) \land (y \mle t) \land \V(t)) \implies (z \mle t)
      }{
          \ownGhost{\gamma}{x}
          \land
          \ownGhost{\gamma}{y}
          \entails
          \ownGhost{\gamma}{z}
      }
    \end{mathpar}
  \end{center}
\end{subfigure}
\hfill
\begin{subfigure}[t]{.36\textwidth}
  \begin{center}\small
    \textbf{Basic Properties of View Shifts}

    \begin{mathpar}
    \inferhref{VS-Hoare}{VS-Hoare}{
      P' \vs[\mask] P \quad
      \hoare{P}{e}{Q}_\mask\quad
      Q \vs[\mask] Q'
    }{
      \hoare{P'}{e}{Q'}_\mask
    }

    \inferhref{VS-Weaken}{VS-Weaken}{
        P \vs[\mask_1] Q
    }{
        P \vs[\mask_1 \cup \mask_2] Q
    }
    \end{mathpar}
  \end{center}
\end{subfigure}
  \vspace{-2mm}
  \caption{\label{fig-pcm-basic-rules} \small
      Deduction rules for PCM-based ghost state and view shifts.
  }
  \vspace{-3mm}
\end{figure}

\paragraph{Using the overlapping conjunction}
We make a point to include a rule for overlapping conjunction, since in dealing with shared state we often have the potential for overlap.
\ruleref{PCM-And} lets us deduce that $\ownGhost{\gamma}{x} \land \ownGhost{\gamma}{y} \entails \ownGhost{\gamma}{z}$
under the premise that, if for all valid states $t$ with $x$ and $y$ as sub-states,
$z$ is also a sub-state.
As an example, consider $\ownGhost{\gamma}{\exConstructor(x)} \land \ownGhost{\gamma}{\exConstructor(y)}$.
If $x \ne y$, then $\ownGhost{\gamma}{\exConstructor(x)} \land \ownGhost{\gamma}{\exConstructor(y)} \entails \ownGhost{\gamma}{\mundef}$  by \ruleref{PCM-And}
because \emph{no} valid state includes both $\ownGhost{\gamma}{\exConstructor(x)}$ and $\ownGhost{\gamma}{\exConstructor(y)}$.
Then we can conclude that $\ownGhost{\gamma}{\exConstructor(x)} \land \ownGhost{\gamma}{\exConstructor(y)} \entails (x = y)$.

\subsection{The Guarding Operator and its Elementary Deduction Rules} \label{sec:logic:main}

The fundamental building block of the {\langname} logic is the $\guards$ operator,
pronounced \emph{guards}.
When we write $G \guards[\mask] I$, we call $G$ the \emph{guard} and $I$
the \emph{guarded proposition} or sometimes the \emph{guarded invariant},
and this means that having $G$ allows shared access to $I$.
Guards, like view shifts, are annotated with a mask $\mask$, as we discuss below.
The basic rules for $\guards[\mask]$ are given in \autoref{fig:guard:logic}.

For example, \ruleref{Guard-Refl} says that a $P$ represents a shared $P$,
while \ruleref{Guard-Trans} says that
if $Q$ is a shared $R$,
then a shared $Q$ is also a shared $R$.
\ruleref{Guard-Pers} and \ruleref{Unguard-Pers} show how persistent propositions can move into or out from under a $\guards$.
\ruleref{Guard-Upd} says that when
an update $P * A \vs P * B$ is valid, then we can perform this update even when
we have a shared $P$.

\begin{figure}
\begin{center}\small
  \textbf{Deduction rules for guarded resources}

  \textbf{Persistent Propositions:}\quad
    $P \guards[\mask] Q$
  \quad
  \text{(where $P, Q: \iProp,~~ \mask: \powset(\Name)$)}
  \begin{mathpar}
    \axiomhref{Guard-Refl}{Guard-Refl}{
      P \guards[\mask] P
    }

    \axiomhref{Guard-Trans}{Guard-Trans}{
      (P \guards[\mask] Q) *
      (Q \guards[\mask] R) \entails
      (P \guards[\mask] R)
    }

    \axiomhref{Guard-Split}{Guard-Split}{
      P * Q \guards[\mask] P
    }

    \axiomhref{Guard-Weaken-Mask}{Guard-Weaken-Mask}{
        (G \guards[\mask_1] P)
        \entails
        (G \guards[\mask_1 \cup \mask_2] P)
    }

    \inferhref{Guard-Pers}{Guard-Pers}{
        \persistent{C}
    }{
        (G \guards[\mask] A) * C \entails (G \guards[\mask] A * C)
    }

    \inferhref{Unguard-Pers}{Unguard-Pers}{
      A * B \entails C \quad
      \persistent{C}
    }{
      G * (G \guards[\mask] A) * B
      \vs[\mask]
      G * (G \guards[\mask] A) * B * C
    }

    \inferhref{Guard-Upd}{Guard-Upd}{
        \mask_1 \cap \mask_2 = \emptyset
    }{
        (G \guards[\mask_1] P)
        *
        (P * A \vs[\mask_2] P * B)
        \entails
        (G * A \vs[\mask_1 \cup \mask_2] G * B)
    }

    \axiomhref{PointProp-Own}{PointProp-Own}{
        \pointprop{\ownGhost{\gamma}{x}}
    }

    \inferhref{Guard-Implies}{Guard-Implies}{
        A \entails P
        \quad
        \pointprop{P}
    }{
      (G \guards[\mask] A)
      \entails
      (G \guards[\mask] P)
    }

    \inferhref{Guard-And}{Guard-And}{
        A \land B \entails P
        \quad
        \pointprop{P}
    }{
      (G \guards[\mask] A)
      *
      (G \guards[\mask] B)
      \entails
      (G \guards[\mask] P)
    }

    \inferhref{PointProp-Sep}{PointProp-Sep}{
        \pointprop{P} \quad \pointprop{Q}
    }{
        \pointprop{P * Q}
    }
  \end{mathpar}
\end{center}
  \vspace{-3mm}
  \caption{\label{fig:guard:logic}
      Deduction rules for $\guards$ introduced by {\langname}.
  }
  \vspace{-3mm}
\end{figure}

\paragraph{Mask Sets}

We use mask sets, $\mask$, to track the ``sources'' of the sharing for both guard relations ($\guards[\mask]$) and view shifts ($\vs[\mask]$).
The sharing sources---as we will see later---are called storage protocols, and each storage protocol has a name $\gname$.
A mask is a set of such names, and it can be considered an over-approximation of the set of storage protocols involved in the derivation of a given relation.
(In \autoref{sec:soundness}, we will see that this interpretation of $\vs[\mask]$ follows
from its usual Iris definition.)

We need to track these because we need to be careful when we apply \ruleref{Guard-Upd}.
If we applied \ruleref{Guard-Upd} twice while ignoring its disjointness condition, we could potentially ``double up'' a proposition that is shared between multiple guard objects.
For example, if we had $G_1 \vs[\mask] P$ and $G_2 \vs[\mask] P$, we could use $G_1 * G_2$ to perform an update that would only be possible if we had $P * P$.
The disjointness condition prevents us from doing this.
Those familiar with Iris may observe that this is similar to invariant reentrancy, so it should be no surprise that we solve the problem the same way, that is, via mask sets.

\paragraph{Guards and implications}
One might expect a rule where we use
$G \guards I$ and $I \entails J$ to conclude $G \guards J$.
This \emph{does} work when we can write $I = J * J'$ for some $J'$ (\ruleref{Guard-Split}),
but it does not hold in general:
consider, for example, a judgment such as
$(\ell \pointsto 1) \entails (\exists x.~ \ell \pointsto x)$.
It would be unsound if a user sharing $\ell \pointsto 1$ could ``downgrade'' it to
the right-hand side; that user could then update $\ell$ to a different value and invalidate
the proposition the other users were relying on.

Interestingly, it turns out that there are some propositions $J$ such that any judgment
$I \entails J$ \emph{can} always be ``split'' into $I = J * J'$.
Specifically, this happens whenever $J$ is of the form $\ownGhost{\gamma}{x}$ or a conjunction thereof.
We call these \emph{point propositions} and indicate them by $\pointprop{J}$ (\ruleref{PointProp-Own}, \ruleref{PointProp-Sep}).
For such $J$, we can indeed conclude $G \guards J$ (\ruleref{Guard-Implies}).

\paragraph{Overlapping Conjunction}
How can we compose shared state?  We certainly cannot have a rule like
$(G \guards[\mask] A) * (H \guards[\mask] B) \entails ((G * H) \guards[\mask] (A * B))$.
After all, $A$ and $B$ might be shared from the same source and thus not be properly separated.
Somewhat surprisingly, this rule is not even sound if we require the masks to be disjoint.
(See \appref{app:counterexample} for a concrete counterexample.)

Instead of using $*$, we use $\land$.
One might instead conjecture a $\land$-based rule like
$(G \guards[\mask] A) * (G \guards[\mask] B) \entails (G \guards[\mask] A \land B)$;
this rule still is not sound on its own (again, see \appref{app:counterexample} for a concrete counterexample),
but fortunately, it becomes sound as long as we add another point proposition condition (\ruleref{Guard-And}).
This rule is especially useful in combination with \ruleref{PCM-And}, which can be used to deduce the premise of \ruleref{Guard-And}.

\subsection{Using $\guards$ in a program logic} \label{sec:program-logic-heap}

\paragraph{Deriving heap rules} 
Iris is not a separation logic for a single programming language; rather, it is a general
separation logic \emph{framework} which can be used to instantiate a program logic
for any user-provided
programming language. In other words, rules like the following, which might be considered
``primitive'' rules within a program logic, can actually be derived soundly within Iris.
{\small\begin{mathpar}
\axiomhref{Heap-Ref}{Logic-Heap-Ref}{
  \inlinehoaresmall{ }~  \langref(v) ~\inlinehoaresmall{ \ell.~  \ell \pointsto v }
}
\hspace{-0.05in}

\hspace{-0.05in}
\axiomhref{Heap-Free}{Logic-Heap-Free}{
  \inlinehoaresmall{ \ell \pointsto v }~  \langfree(v) ~\inlinehoaresmall{ }
}
\hspace{-0.05in}

\hspace{-0.05in}
\axiomhref{Heap-Write}{Logic-Heap-Write}{
  \inlinehoare{ \ell \pointsto v }~ \ell ~\langassign~ v' ~\inlinehoare{ \ell \pointsto v' }
}
\hspace{-0.05in}

\hspace{-0.05in}
\axiomhref{Heap-Read}{Logic-Heap-Read}{
  \inlinehoaresmall{ \ell \pointsto v }~  \langbang\ell ~\inlinehoaresmall{ r.~  \ell \pointsto v * v = r }
}
\end{mathpar}}%
Let us overview this process, and then
explain how it works with {\langname}'s $\guards$ in the picture.

To instantiate a program logic, the user provides their programming language and its operational
semantics. Here, we consider heap semantics operating over a state
given by $\sigma : \Loc \fpfn \Value$, with allocation ($\langref$),
deallocation ($\langfree$), assignment ($\langassign$) and reading ($\langbang$).
Next, the user gives meaning to the heap state $\sigma$ within the separation logic
by defining an interpretation of the heap state as a proposition, $\HeapInterp(\sigma) : \iProp$,
along with propositions to be manipulated by the user within the program logic (here, $\ell \pointsto v$).
Finally, they prove the primitive heap rules via corresponding updates or entailments.
For example, the following suffice to show the above four Hoare rules.
{\small\begin{align*}
  \HeapInterp(\sigma) &\vs \exists \ell.~ \HeapInterp(\sigma[\ell := v]) * (\ell \pointsto v')
  & \rulelabel{AllocUpd}
\\
  \HeapInterp(\sigma) * (\ell \pointsto v) &\vs \HeapInterp(\sigma\backslash\{\ell\})
  & \rulelabel{FreeUpd}
\\
  \HeapInterp(\sigma) * (\ell \pointsto v) &~\entails~ (\sigma(\ell) = v)
  & \rulelabel{ReadEq}
\\
  \HeapInterp(\sigma) * (\ell \pointsto v) &\vs \HeapInterp(\sigma[\ell := v']) * (\ell \pointsto v')
  & \rulelabel{WriteUpd}
\end{align*}}%
Thus, it suffices for the user to
construct $\HeapInterp(\sigma)$ and $\ell \pointsto v$ so that the above hold;
this can be done via a custom PCM construction, using
\ruleref{PCM-Valid} to prove \ruleref{ReadEq},
\ruleref{PCM-Update} to prove \ruleref{WriteUpd} and \ruleref{FreeUpd},
and \ruleref{PCM-Alloc} to prove \ruleref{AllocUpd}.

Now, the new rule we want to construct is, for any $\ell, v, \mask, \mask_1$,
{\small\begin{mathpar}
    \axiomhref{Heap-Read-Shared}{LogicSec-Heap-Read-Shared}{
      \hoareShared[\ell \pointsto v][\mask]~
      \{  \} ~ \langbang\ell ~ \{ v'.~ v = v' \}_{\mask \cup \mask_1}
    }
\end{mathpar}}%
Expanding the notation, this is equivalent to, for any $G: \iProp$,
{\small\begin{align*}
    \{ G * (G \guards[\mask] {\colorShared (\ell \pointsto v)}) \} ~ \langbang\ell ~ \{ v'.~ G * (v = v') \}_{\mask \cup \mask_1}
\end{align*}}%
This follows from,
{\small\begin{align*}
    (G \guards[\mask] {\colorShared (\ell \pointsto v)}) ~~\entails~~
    \HeapInterp(\sigma) * G
    \vs[\mask \cup \mask_1]
    \HeapInterp(\sigma) * G * (\sigma(\ell) = v)
\end{align*}}%
and this in turn follows from \ruleref{Unguard-Pers} and \ruleref{ReadEq}.
Notably, we do not need to re-do the construction of $\HeapInterp(\sigma)$
or $\ell \pointsto v$ to support the derivation of
\ruleref{LogicSec-Heap-Read-Shared}.
Along with the new deduction rules for $\guards$, the old construction ``just works.''

\paragraph{Atomic Invariants}

Propositions shared via guarding can serve as \emph{atomic invariants}; i.e., we can
obtain \emph{exclusive} ownership of a shared proposition for the duration of an atomic
operation, as long as we restore the invariant at the end of the operation.
{\small\begin{mathpar}
    \inferhref{Guard-Atomic-Inv}{Guard-Atomic-Inv}{
      \hoareShared[ \cdots ]~\{ P * X \}~e~\{ Q * X \}_{\mask_1}
      \quad
      \mask \cap \mask_1 = \emptyset
      \quad
      \text{$e$ is atomic}
    }{
      \hoareShared[ X ][\mask]~\hoareShared[ \cdots ]~\{ P \}~e~\{ Q \}_{\mask \cup \mask_1}
    }
\end{mathpar}}%

\paragraph{Non-Atomic Memory}

The heap semantics in the preceding example use sequentially consistent, atomic
heap operations. But what about other memory ordering models?

We can also apply {\langname} to heap semantics that model \emph{non-atomic memory access}, i.e.,
memory accesses for which data races are entirely disallowed, alongside atomic operations.
Non-atomic memory has been modeled before in Iris, e.g., by RustBelt~\cite{rustbelt},
which models each non-atomic operation as two execution steps in order to detect overlapping
operations.
We can apply {\langname} to this situation, and prove \ruleref{LogicSec-Heap-Read-Shared}
for non-atomic reads;
however, the proof is slightly more challenging
than it is for the purely-atomic heap semantics, primarily because the heap semantics
have to model non-atomic reads as effectful operations.
To get around this, we need to be slightly clever in our definition of $\HeapInterp(\sigma)$;
see \appref{app:nonatomic} for a sketch.

\subsection{Storage Protocols} \label{sec:protocol}

\begin{figure}
  \begin{subfigure}[t]{.48\textwidth}
  \begin{center}\small
  A \textbf{storage protocol} consists of:

  A \emph{storage monoid}, that is, a partial commutative monoid $(S, \cdot, \V)$, where, 
  \begin{align*}
      \forall a &.~ a \cdot \uunit = a\\
      \forall a, b &.~ a \cdot b = b \cdot a\\
      \forall a, b, c &.~ (a \cdot b) \cdot c = a \cdot (b \cdot c)\\
      &\phantom{.}~ \V(\uunit)\\
      \forall a, b &.~ a \mle b \land \V(b) \implies \V(a)
  \end{align*}

  A \emph{protocol monoid}, that is, a (total) commutative monoid $(P, \cdot)$, 
  with an arbitrary predicate $\Inv : P \to \bool$ and function $\Sfunc : P|_{\Inv} \to S$
  (i.e., the domain of $\Sfunc$ is restricted to the subset of $P$ where $\Inv$ holds)
  where,
  \begin{align*}
      \forall a &.~ a \cdot \uunit = a\\
      \forall a, b &.~ a \cdot b = b \cdot a\\
      \forall a, b, c &.~ (a \cdot b) \cdot c = a \cdot (b \cdot c)\\
      \forall a &.~ \Inv(a) \implies \V(\Sfunc(a))
  \end{align*}
  Note that $\Inv$ (unlike $\V$) is \emph{not} necessarily closed under $\mle$.
  ~\\
  \textbf{Derived relations} for storage protocols:\\
  ~\\
  For $p, p' : P$ and $s, s' : S$, define:
  \begin{align*}
      (p, s) \exchange (p', s') &\eqdef
          \forall q.~ \Inv(p \cdot q) \implies (\Inv(p' \cdot q)\\
            &
              \land \V(\Sfunc(p \cdot q) \cdot s)\\
            & \land \Sfunc(p \cdot q) \cdot s
                  = \Sfunc(p' \cdot q) \cdot s')\\
      p \withdraws (p', s') &\eqdef (p, \uunit) \exchange (p', s')\\
      (p, s) \deposits p'\phantom{(,s')} &\eqdef (p, s) \exchange (p', \uunit)\\
      p \transitions p'\phantom{(,s')} &\eqdef (p, \uunit) \exchange (p', \uunit)\\
      p \guardsFO s\phantom{(,p')} &\eqdef 
        \forall q.~ \Inv(p \cdot q) \implies s \mle \Sfunc(p \cdot q)
  \end{align*}
  \end{center}
  \vspace{-2mm}
  \caption{ \label{fig:protocol:definition}
    Definitions.
  }
  \end{subfigure}
  \hfill
  \begin{subfigure}[t]{.48\textwidth}
  \begin{center}\small
    \textbf{Storage Protocol Logic}

    Instantiated for a given storage protocol $(S, \cdot, \V), (P, \cdot), \Inv, \Sfunc$

    \textbf{Propositions:}\quad
      $\protOwn{p}{\gamma}$

    \textbf{Persistent propositions:}\quad
      $\maps(\gamma, F)$

    \text{(where $\gamma: \Name,~~ F: S \to \iProp,~~ p: P$)}

    \begin{mathpar}
      \stacktwo{
      \RespectsComposition(F) ~~\eqdef~~
            \text{($F(\uunit) \bothentails \true$)\phantom{xxxxxx}}}
      {\text{and $\forall x,y.~ \V(x \cdot y) \implies (F(x \cdot y) \bothentails F(x) * F(y))$}}

      \inferhref{SP-Alloc}{SP-Alloc}{
          \RespectsComposition(F)
          \quad
          \Inv(p)
          \quad
          \text{$\mathcal{N}$ infinite}
      }{
          F(\Sfunc(p))
          \vs
          \exists \gamma.~ \maps(\gamma, F) * \protOwn{p}{\gamma} * (\gamma \in \mathcal{N})
      }

      \inferhref{SP-Exchange}{SP-Exchange}{
          (p,s) \exchange (p', s')
      }{
          \maps(\gamma, F)\entails
          (\later F(s)) * \protOwn{p}{\gamma}
          \vs[\gamma]
          (\later F(s')) * \protOwn{p'}{\gamma}
      }

      \inferhref{SP-Deposit}{SP-Deposit}{
          (p,s) \deposits p'
      }{
          \maps(\gamma, F)\entails
          (\later F(s)) * \protOwn{p}{\gamma}
          \vs[\gamma]
          \protOwn{p'}{\gamma}
      }

      \inferhref{SP-Withdraw}{SP-Withdraw}{
          p \withdraws (p', s')
      }{
          \maps(\gamma, F)\entails
          \protOwn{p}{\gamma}
          \vs[\gamma]
          (\later F(s')) * \protOwn{p'}{\gamma}
      }

      \inferhref{SP-Update}{SP-Update}{
          p \transitions p'
      }{
          \maps(\gamma, F)\entails
          \protOwn{p}{\gamma}
          \vs[\gamma]
          \protOwn{p'}{\gamma}
      }

      \axiomhref{SP-PointProp}{SP-PointProp}{
          \pointprop{\protOwn{p}{\gamma}}
      }

      \inferhref{SP-Guard}{SP-Guard}{
          p \guardsFO s
      }{
          \maps(\gamma, F)\entails
          \protOwn{p}{\gamma}
          \guards[\gamma]
          (\later F(s))
      }

      \axiomhref{SP-Unit}{SP-Unit}{
          \maps(\gamma, F) \entails \protOwn{\uunit}{\gamma}
      }

      \axiomhref{SP-Sep}{SP-Sep}
      {
          \protOwn{p}{\gamma} *
          \protOwn{q}{\gamma}
          \bothentails
          \protOwn{p \cdot q}{\gamma}
      }

      \axiomhref{SP-Valid}{SP-Valid}{
          \protOwn{p}{\gamma} \entails
              \exists q.~ \Inv(p \cdot q)
      }
    \end{mathpar}
  \end{center}
  \vspace{-2mm}
  \caption{ \label{fig:protocol:logic}
      Deduction rules.
  }
  \vspace{-3mm}
  \end{subfigure}
  \caption{Storage protocols and derived relations.
  }
  \vspace{-3mm}
\end{figure}

{\langname}'s \emph{storage protocol} is a formulation of custom ghost state whose unique feature is its laws allowing deductions of nontrivial $\guards$ propositions.
Storage protocols are similar to the ghost state presented earlier, which embeds elements of a monoid as separation logic propositions $\ownGhost{\gamma}{x}$
and uses a derived relation ($\transitions$) to deduce updates ($\vs)$.
The storage protocol formulation, however, is given 
as a relationship between \emph{two} monoids, a \emph{protocol monoid} $P$ and a \emph{storage monoid} $S$.
Elements $p : P$ are embedded as propositions $\protOwn{p}{\gamma}$, analogously to $\ownGhost{\gamma}{x}$, while
elements $s : S$ are embedded via an arbitrary proposition family $F : S \to \iProp$, which
is specified upon initialization of the ghost state (\ruleref{SP-Alloc}).
The update relation and the resulting $\vs$ propositions are complicated by the need to account for $S$,
and we also include a new derived relation, $\guardsFO$, from which we deduce guards of the form $\protOwn{p}{\gamma} \guards[\mask] F(s)$.
\autoref{fig:protocol:definition} gives the precise definitions for these derived relations.

Let us first discuss the role of $S$, and its impact on our definition of updates.
$P$ and $S$ are related by a \emph{storage function} $\Sfunc : P \to S$, which
intuitively associates to each possible state of $P$ some value $s : S$ that is ``stored,''
and since the stored value might change upon any update, we need to account for this in the definition of updates.
For example,
when the $P$-state updates, the corresponding stored state ($\Sfunc(p)$) might also change; if
the stored state changes from $t$ to $t\cdot s$, we consider $s$ to be ``deposited'';
if the state changes the other way, we consider it withdrawn.
The most general form of an exchange is given by $\exchange$, with the other three as special cases:
A withdraw ($\withdraws$) occurs when ``nothing'' (i.e., the unit) is deposited,
a deposit ($\deposits$) occurs when nothing is withdrawn,
and a ``normal'' update ($\transitions$) occurs when the stored value remains constant.
These relations then give rise to updates in the separation logic
(\ruleref{SP-Exchange}, and its special cases,
\ruleref{SP-Deposit},
\ruleref{SP-Withdraw},
\ruleref{SP-Update}), operating on $\protOwn{p}{\gamma}$ and $F(s)$.
Specifically, a deposit of $s$ allows an update that gives up ownership of $F(s)$,
while a withdraw of $s$ allows an update that obtains ownership of $F(s)$.

Now, with the ability to ``deposit'' elements into the protocol and ``withdraw'' them,
we can add the ability to have shared access to those stored elements:
the derived relation $p \guardsFO s$ gives rise to $\protOwn{p}{\gamma} \guards F(s)$ by \ruleref{SP-Guard}.
Let us unpack the definition of $\guardsFO$ to understand intuitively why this should work:
$p \guardsFO s$ is defined as $\forall q.~ \Inv(p \cdot q) \implies s \mle \Sfunc(p \cdot q)$;
this essentially says that $p$ is a ``witness'' that the value $s$ is stored;
i.e., if we have ownership of $p$, then any ``completion'' of the state $p \cdot q$, which is valid according to the validity function $\mathcal{C}$,
must be storing something $\mge s$.

\emph{Initialization of a protocol.}
When initializing a protocol~(\ruleref{SP-Alloc}) we get to specify $F$,
and we also get to specify the initial
element $s : S$ while giving up $F(s)$, the initial proposition to be stored.
Inheriting yet another trick from Iris,
we can also specify a \emph{namespace} $\mathcal{N}$, a subset of $\Name$,
that the resulting protocol name has to be in.
Using fixed namespaces (such as $\NamespaceFrac$ or $\NamespaceCount$ from
\autoref{fig:frac:count:logic}) is often more convenient than managing individual
names $\gamma$ that cannot be known \emph{a priori}.

\begin{example}[Fractional protocol for a single proposition] \label{ex:frac}
  To start simple, let us suppose we have a single proposition, $Q$, we would like to manage.
  Set the protocol monoid $P \eqdef \mathbb{Q}_{\ge 0}$ and the storage monoid $S \eqdef \mathbb{N}$,
  with composition as addition in both cases and a unit of $0$.
  Set $\Inv$ to be true exactly on the integers, and for integers $n$, set $\Sfunc(n) \eqdef n$.
  Let $\V$ always be true. Now, we have
  the exchange
  $(1, 0) \exchange (0, 1)$ (also written as a withdraw, $1 \withdraws (0, 1)$)
  and the reverse,
  $(0, 1) \exchange (1, 0)$ (also written as a deposit, $(0, 1) \deposits 1$).
  Finally, for any $q > 0$, we have $q \guardsFO 1$.
  This is the key property that says the fraction $q$ can act as a read-only element,
  and it follows from the following argument: if $q' \ge q$ and $\Inv(q')$ holds, then
  $q'$ is an integer and $\Sfunc(q') = q' \ge 1$.

  Finally, set the proposition family $F(n) \eqdef \MediumAsterisk^n ~ Q$, i.e., $Q$ conjoined
  $n$ times. Now we can say that,
  {\small\begin{align*}
    \true &\vs \exists \gamma.~ \maps(\gamma, F) && \text{(via \ruleref{SP-Alloc})}\\
    \maps(\gamma, F) \entails \later Q &\vs[\gamma] \protOwn{1}{\gamma} && \text{(via \ruleref{SP-Deposit})}\\
    \maps(\gamma, F) \entails \protOwn{1}{\gamma} &\vs[\gamma] \later Q && \text{(via \ruleref{SP-Withdraw})}\\
    \maps(\gamma, F) \entails \protOwn{q}{\gamma} &\guards[\gamma] \later Q
        ~\text{\phantom{aaaaaaa} (for $q > 0$)} && \text{(via \ruleref{SP-Guard})}
  \end{align*}}%
\end{example}

\begin{example}[Fractional memory permissions] \label{ex:frac:mem}
  Assume points-to propositions $\ell \pointsto v$ are given.
  We wish to construct $\ell \pointstofrac{q} v$ and the laws as given in
  \autoref{fig:frac:count:logic}.

  We take $P = (\Loc \times \Value) \fpfn \mathbb{Q}_{\ge 0}$ and
  $S = (\Loc \times \Value) \fpfn \mathbb{N}$, defining $\cdot, \V, \Inv, \Sfunc$
  elementwise using the definitions of the previous example.
  Define $F$ such that $F([(\ell,v) \mapsto 1]) = \ell \pointsto v$.
  Instantiate the protocol to obtain a location $\gamma \in \NamespaceFrac$, and then set
  $\ell \pointstofrac{q} v \eqdef \protOwn{[(\ell, v) \mapsto q]}{\gamma}$.
  From here we can derive the appropriate withdraw, deposit, and guard.
\end{example}

The counting protocol (\autoref{fig:frac:count:logic}) can be done similarly.
(See \appref{app:count} for details.)

\begin{example}[Forever Protocol] \label{ex:forever}
The most basic sharing pattern is to make something freely shareable forever
(analogous to Iris invariants $\knowInv{}{Q}$).
We can express this succinctly by guarding $Q$ with $\true$:
$Q \vs (\true \guards[\NamespaceForever] \later Q)$.
To derive this, we use a storage protocol: let the protocol monoid
$P$ be the trivial monoid $\{\uunit\}$ and the storage monoid $S$ be $\Ex(\UnitType)$,
with $\Sfunc(\uunit) \eqdef \exConstructor(1)$.
This protocol has no interesting updates; it can only be initialized.
Let $F(\exConstructor(1)) \eqdef Q$.
By \ruleref{SP-Alloc} we have $Q \vs \exists \gamma.~ \maps(\gamma, F) * \protOwn{\uunit}{\gamma} * (\gamma \in \NamespaceForever)$.
Now we can chain
$\true \guards \protOwn{\uunit}{\gamma}$ (by \ruleref{Guard-Pers})
and
$\protOwn{\uunit}{\gamma} \guards[\NamespaceForever] (\later Q)$ (by \ruleref{SP-Guard}).
\end{example}

\subsection{Handling the later modality $\later$} \label{sec:logic:later}

Note that some rules in \autoref{fig:protocol:logic} use the \emph{later modality}, $\later$,
a feature of step-indexed logics like Iris.
In {\langname}, $\later$ allows us to dynamically
specify the proposition families $F$ during protocol initialization (\ruleref{SP-Alloc});
without $\later$, this would be unsound.
If we gave up that ability and instead specified all families \emph{a priori}, we could remove
$\later$ from the rest of the rules. (This is analogous to Iris requiring $\later$ for dynamically allocated invariants.)
{\langname} provides rules to eliminate $\later$ from within guards:
{\small\begin{mathpar}
  \inferhrefright{Later-Guard}{Later-Guard}{
      \timeless{P}
  }{
      (\later P) \guards[\mask] P
  }

  \inferhrefright{Later-Pers-Guard}{Later-Pers-Guard}{
      \timeless{P} \quad \persistent{Q}
  }{
      G * (G \guards[\mask] \later (P * Q))
      \vs[\mask]
      G * \later (G \guards[\mask] (P * Q))
  }
\end{mathpar}}%
\emph{Timelessness}~\cite{iris3} is a technical condition that effectively says a proposition is ``independent of the step-index,'' which makes it easier to account for $\later$ modalities.
Timelessness holds for both the PCM-based $\ownGhost{\gamma}{x}$ and our $\protOwn{x}{\gamma}$ propositions. (Technically, this is because PCMs are special cases of ``discrete CMRAs.'')
However, $\maps(\gamma, F)$ is \emph{not} timeless, since it depends recursively on $\iProp$,
but since it \emph{is} persistent, we can use \ruleref{Later-Pers-Guard} for such propositions.

\section{RwLock Example: Verifying a Custom Protocol for Sharing State} \label{sec:rwlock}

\begin{figure}
\begin{subfigure}[t]{.48\textwidth}
\begin{align*}\small\begin{array}{l}
\rwlocknew() \eqdef
  \{ \exc: \langref(\false), \rc: \langref(0) \}\\
  \\
\acquireExc(\rw) \eqdef\\
  \langindent \langdo \\
  \langindent \langindent \langlet ~ \textit{success} = \CAS(\rw.\exc, \false, \true)\\
  \langindent \languntil ~ \textit{success}\\
  \langindent \langdo \\
  \langindent \langindent \langlet ~ r =~ \langbang \rw.\rc\\
  \langindent \languntil ~ r = 0\\
  \\
\releaseExc(\rw) \eqdef
  \rw.\exc ~\langassign~ 0
\end{array}\end{align*}
\end{subfigure}
\hfill
\begin{subfigure}[t]{.48\textwidth}
\begin{align*}\small\begin{array}{l}
\rwlockfree(\rw) \eqdef
  \langfree(\rw.\exc); \langfree(\rw.\rc)\\
  \\
\acquireShared(\rw) \eqdef\\
  \langindent \langdo \\
  \langindent \langindent \FetchAdd(\rw.\rc, 1);\\
  \langindent \langindent \langlet ~ \exc =~ \langbang \rw.\exc ~ \langin \\
  \langindent \langindent \langif ~ \exc ~ \langthen ~ \FetchAdd(\rw.\rc, -1);\\
  \langindent \languntil ~ \exc = \false\\
  \\
\releaseShared(\rw) \eqdef
  \FetchAdd(\rw.\rc, -1)
\end{array}\end{align*}
\end{subfigure}
\caption{\label{fig:rwlock:impl}
  Implementation of a reader-writer lock. 
  We assume all heap operations are atomic, including $\CAS$ (compare-and-swap) and $\FetchAdd$.
}
\end{figure}

\begin{figure}
    \begin{center}
        \includegraphics[width=\textwidth]{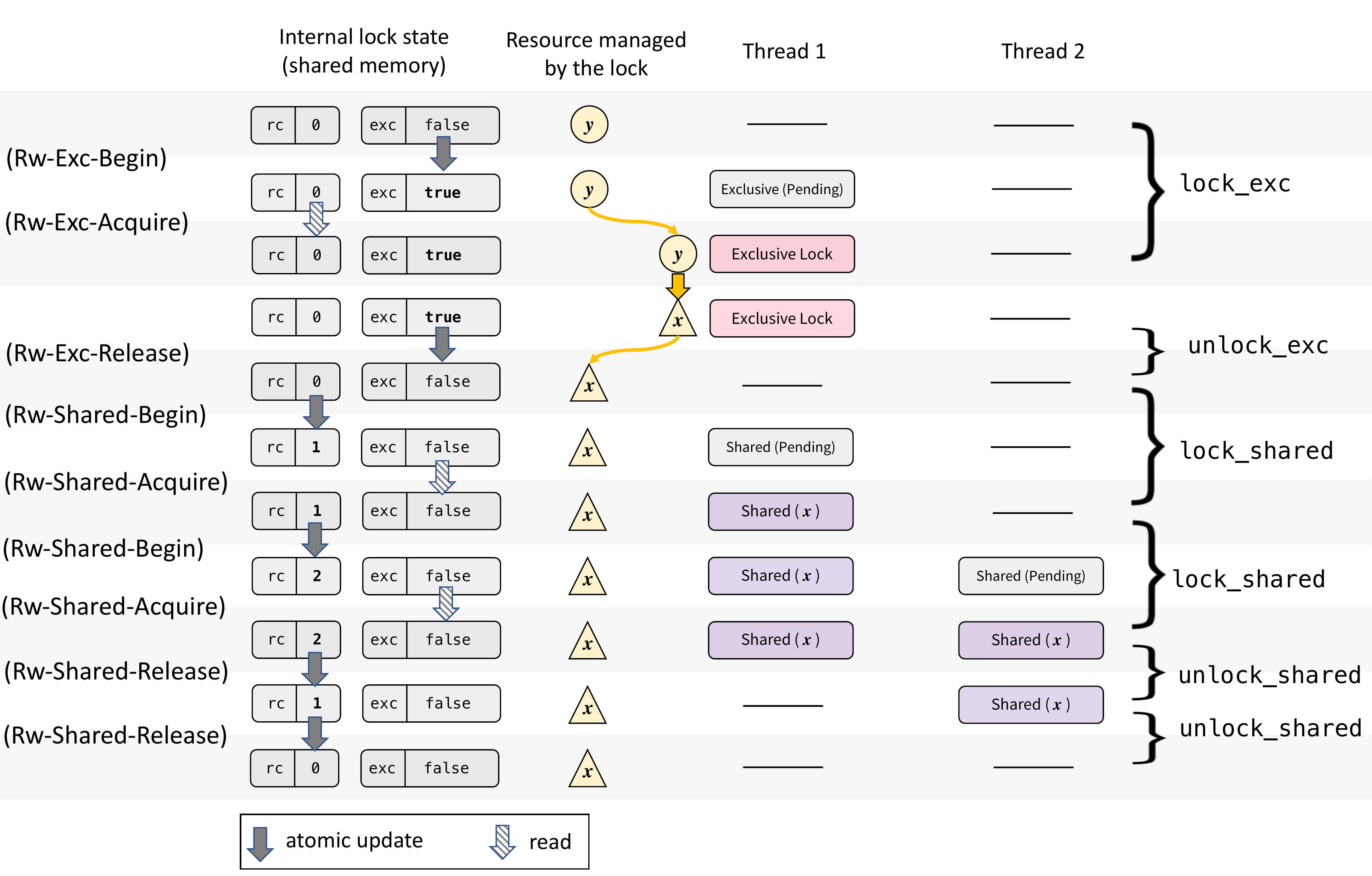}
    \end{center}
    \caption{
        \label{fig:execution}
        Example execution of two threads using a shared reader-writer lock.
        This is an ``ideal'' execution, without contention or retries.
        First, we see Thread 1 acquire exclusive lock. This gives them exclusive
        control over the resource, which they can therefore modify (here, changing it
        from $y$ to $x$) before releasing the lock.
        We then see both threads acquire a shared lock, where they simultaneously
        have read access to the $x$ resource.
        On the left, we annotate each step with the ghost resource update from \autoref{fig:rwlock:logic} it corresponds to.
    }
\end{figure}

Our two case studies are arranged to show the two ``halves'' of Leaf:
the first one (this section) demonstrates the verification
of a sharing protocol that lets the client acquire shared state,
while our second one (\autoref{sec:ht}) shows how a client can make use of the shared
state.

Specifically, in this section, we verify a reader-writer lock,
one which is slightly
more complicated than that which is captured directly by a standard permission logic,
a situation which the storage protocol was designed for.
The implementation of our reader-writer lock in shown in
\autoref{fig:rwlock:impl}, with an example execution trace in \autoref{fig:execution}.
The implementation's main complication here is the fact that acquiring a lock is a two-step process: a thread might increment the reference counter, but then fail to acquire the lock in the second step. Hence, the physical value of the reference counter may not match the number of extant read-references.

Initially, this design might seem strange---why not just put all the data in a single atomic field
to simplify the design?
However, the use of distinct fields is an essential element of more complex lock designs, such as the multi-counter design mentioned in the introduction, where each counter goes on a different cache line.  In fact, simply having an intermediate state at all captures most of the complexity of the multi-counter design, so we use the single-counter implementation here---see
\appref{app:rwmulti} for details on the multi-counter case.

\paragraph{Proof Overview}
We tackle the proof in two stages: first, we devise some useful ghost resources;
then, we use those resources in the program logic to prove
the implementation meets the specification (\autoref{fig:rwlock:spec}).
The key is to find the right resources and their relationships that we need.

The RwLock specification~(\autoref{fig:rwlock:spec}) already indicates three propositions that we need to construct
in one way or another:
  $\IsRwLock(\rw, \gamma, F)$,
  $\ExcGuard(\gamma)$, and $\SharedGuard(\gamma, x)$.
  We also have $\SharedGuard(\gamma, x) \guards[\gamma] F(x)$ as a desired property.
  This gives us a reason to use a storage protocol: it allows us to construct new resources
  and derive $\guards$ relationships.
  Since the storage protocol and its resulting resources are more elaborate than in the
  previous examples, we will take the time here to explain exactly how to come up with the protocol.

  First, we naturally need a component to represent the lock's internal state,
  which we call $\Fields(\gamma, \exc, \rc, x)$, containing both
  the $\exc$ and $\rc$ fields, and also the stored value, $x$.
  We can tie the first two fields to the physical, in-memory values with a proposition
  like the following:
  \[
      \IsRwLock(\rw, \gamma, F) \eqdef 
          \exists \exc,\rc,x .~
            \Fields(\gamma, \exc, \rc, x)
            * (\rw.\exc \pointsto \exc)
            * (\rw.\rc \pointsto \rc)
            * \ldots
  \]
  Next, we use resources to represent the intermediate states that occur
  during lock acquisition. For example, write-lock acquisition
  has a moment where we have set $\exc$ but not observed $\rc$; likewise,
  read-lock acquisition has a temporary state where we have incremented $\rc$
  but not observed $\exc$.
  We use $\ExcPending(\gamma)$ and $\SharedPending(\gamma)$ to represent these
  states.

  So for example, to prove $\acquireExc$, which has the intended specification,
{\small\begin{align*}
    \hoareShared[ \IsRwLock(\rw, \gamma, F) ]~
    \{ \}~\acquireExc(\rw)~\{ \ExcGuard(\gamma) * \exists x .~ \later F(x) \}
\end{align*}}%
its proof outline should look something like,
{\small\begin{align*}
&\colorShared
[ \exists \exc,\rc,x .~ \Fields(\gamma, \exc, \rc, x)
    * (\rw.\exc \pointsto \exc)
    * (\rw.\rc \pointsto \rc)
 ]\\
&
\colorShared \left\lfloor \color{black}
\begin{array}{l}
  \{ ~ \}\\
  \langdo \\
  \langindent \langlet ~ \textit{success} = \CAS(\rw.\exc, \false, \true)\\
  \languntil ~ \textit{success}\\
  \{ \ExcPending(\gamma) \}\\
  \langdo \\
  \langindent \langlet ~ r =~ \langbang \rw.\rc\\
  \languntil ~ r = 0\\
  \{ \exists x.~ \ExcGuard(\gamma) * \later F(x) \}
\end{array}\right.
\end{align*}}%
With shared access to $\colorShared\rw.\exc \pointsto \exc$
and $\colorShared \rw.\rc \pointsto \rc$, we can use \ruleref{Guard-Atomic-Inv} to perform the requisite atomic $\CAS$ and atomic load.
However, because all these resources are shared, we cannot hold onto them for the duration spanning
both operations at once. Therefore, when performing the $\CAS$, the triple $(\exc,\rc,x)$ used might
be different than the triple used for the later load instruction. 
This is why we cannot track the intermediate ``pending'' state as part of the $\Fields$ resource,
and need to use a separate $\ExcPending$ resource for the thread.

With this sketch in place, we can observe some of the operations we need:
for the $\CAS$ operation, we update $\exc$ from $\false$ to $\true$ and should somehow
obtain $\ExcPending(\gamma)$ in the process. So we need:
{\small\begin{align*}
    \Fields(\gamma, \false, \rc, x) \vs \Fields(\gamma, \true, \rc, x) * \ExcPending(\gamma)
\end{align*}}%
For the second step, reading the $\rc$ value, we find we need:
{\small\begin{align*}
    \Fields(\gamma, \exc, 0, x) * \ExcPending(\gamma)
      \vs
    \Fields(\gamma, \exc, 0, x) * \ExcGuard(\gamma) * \later F(x)
\end{align*}}%
This update requires us to observe that $\rc = 0$, though it does not change $\rc$ or
any of the other fields. It does, however, move us from the pending-exclusive state to the actual
exclusive-lock state, while also acquiring exclusive ownership of the protected resource,
as was our goal.

\autoref{fig:rwlock:logic}  shows all of the operations that we need, including those
we could determine from a similar analysis of the $\acquireShared$ implementation.
This also includes an additional proposition, $\RwFamily(\gamma, F)$, that ties
$\gamma$ to the proposition family $F$.

Now, we just need to use a storage protocol to construct the resources of 
\autoref{fig:rwlock:logic} and prove these the desired updates. Then we can complete the Hoare proofs based on the above plan.

\paragraph{Step 1: Constructing the ghost resources via a storage protocol}
  The first step in building a storage protocol is to determine the storage monoid and the protocol monoid.
  In our example, the storage monoid $S$ can be $\Ex(X)$; i.e., there is either one thing stored,
  or there is not.

  Our primary effort, then, is the protocol monoid $P$. We define $P$ to have a component
  for each class of proposition it needs to support.
  First up is the $\Fields$ proposition, and we know there should always be one such; therefore, we can represent it with $\Ex$.
  Next, there should always be at most one of $\ExcPending$ or $\ExcGuard$, so we can use $\Ex$ for these as well.
  Meanwhile, there might be any number of $\SharedPending$ propositions at a given time, so we can use
  $\mathbb{N}$ for these.

\begin{figure*}
\begin{subfigure}{\textwidth}
\begin{center}\small
\textbf{RwLock Storage Protocol Resource}

\textbf{Propositions:}\quad
    $\Fields(\gamma,\exc,\rc,x)$
    \quad
    $\ExcPending(\gamma)$
    \quad
    $\ExcGuard(\gamma)$
    \quad
    $\SharedPending(\gamma)$
    \quad
    $\SharedGuard(\gamma, x)$

\textbf{Persistent Propositions:}\quad
    $\RwFamily(\gamma, F)$

  \text{(where
  $\gamma : \Name,~~
  \exc : \bool,~~
  \rc : \mathbb{Z},~~
  X : \Set,~~
  x : X,~~
  F : X \to \iProp$)}
\end{center}
    \small\begin{align*}
    F(x) &\vs
    \exists \gamma .~ \Fields(\gamma,\false,0,x)
        * \RwFamily(\gamma, F)
    &\rulelabel{Rw-Init}\\
    \vspace{0.2em}
    \RwFamily(\gamma, F) \entails \hspace{0.5in} \\
    \Fields(\gamma, \false, \rc, x) &\vs[\gamma] \Fields(\gamma, \true, \rc, x) * \ExcPending(\gamma)
    & \rulelabel{Rw-Exc-Begin}\\
    \Fields(\gamma, \exc, 0, x) * \ExcPending(\gamma)
        &\vs[\gamma]
        \Fields(\gamma, \exc, 0, x) * \ExcGuard(\gamma) * \later F(x)
    & \rulelabel{Rw-Exc-Acquire}\\
    \Fields(\gamma, \exc, \rc, y) * \ExcGuard(\gamma) * \later F(x)
        &\vs[\gamma]
            \Fields(\gamma, \false, \rc, x) 
    &\rulelabel{Rw-Exc-Release}\\
    \Fields(\gamma, \exc, \rc, x) &\vs[\gamma]
      \Fields(\gamma, \exc, \rc+1, x) * \SharedPending(\gamma) 
    &\rulelabel{Rw-Shared-Begin}\\
    \Fields(\gamma, \false, \rc, x) * \SharedPending(\gamma)
        &\vs[\gamma]
          \Fields(\gamma, \false, \rc, x) * \SharedGuard(\gamma, x)
    &\rulelabel{Rw-Shared-Acquire}\\
    \Fields(\gamma, \exc, \rc, x) * \SharedGuard(\gamma, y)
        &\vs[\gamma]
            \Fields(\gamma, \exc, \rc - 1, x)
    &\rulelabel{Rw-Shared-Release}\\
    \Fields(\gamma, \exc, \rc, x)  * \SharedPending(\gamma) &\vs[\gamma]
      \Fields(\gamma, \exc, \rc-1, x) 
    &\rulelabel{Rw-Shared-Retry}\\
    \SharedGuard(\gamma, x) &\guards[\gamma] \later F(x)
    &\rulelabel{Rw-Shared-Guard}
    \end{align*}
\end{subfigure}
\caption{ \label{fig:rwlock:logic}
    A custom resource derived via the storage protocol mechanism,
    designed for our particular implementation.
}
\end{figure*}

\newpage
  Finally, for the $\SharedGuard$ propositions, we can have any number,
  but they need to agree on the value of $x$. For this, we use a monoid $\AgN(X)$, which tracks a single value and a count. It is given by,
  \smallskip
  {\small\begin{align*}\begin{array}{c}
      \uunit ~~|~~ \agnConstructor(x, n) ~~|~~ \mundef
      \quad\quad
      \text{where $x : X$ and $n : \mathbb{N}$, $n \ge 1$}\\
      \text{with 
        $\agnConstructor(x, n) \cdot \agnConstructor(x, m) = \agnConstructor(x, n+m)$
        and for $x \ne y$,
        $\agnConstructor(x, n) \cdot \agnConstructor(y, m) = \mundef$
      }
  \end{array}\end{align*}}%
  \smallskip
  
  \noindent All in all, we can now declare our protocol monoid and name its important elements:
  \smallskip
  {\small\[
      P \eqdef
          \Ex(\bool \times \mathbb{Z} \times X)
          \times
          \Ex(\UnitType)
          \times
          \Ex(\UnitType)
          \times
          \mathbb{N}
          \times
          \AgN(X)
  \]}%
  {\small\begin{align*}\begin{array}{rlllll}
      \elemFields(\exc,\rc,x) &\eqdef (\exConstructor((\exc,\rc,x)),& \uunit,& \uunit,&  0,& \uunit)\\
      \elemExcPending &\eqdef (\uunit,& \exConstructor,& \uunit,&  0,& \uunit)\\
      \elemExcGuard &\eqdef (\uunit,&\uunit,&  \exConstructor,& 0,& \uunit)\\
      \elemSharedPending &\eqdef (\uunit,& \uunit,&  \uunit,&  1,& \uunit)\\
      \elemSharedGuard(x) &\eqdef (\uunit,& \uunit,&  \uunit,&  0,& \agnConstructor(x, 1))
  \end{array}\end{align*}}%
  \smallskip

  Now, we need to define $\Sfunc$ and $\Inv$.
  First, $\Sfunc$ determines the element stored; this is given by $x$ in the $\elemFields$ state,
  unless the lock is currently exclusively taken, in which case the storage is empty:
  \smallskip
  {\small
      $$\Sfunc(\exConstructor((\exc,\rc,x)),\_,\uunit,\_,\_) \eqdef \exConstructor(x) \quad\quad
      \Sfunc(\exConstructor((\exc,\rc,x)),\_,\exConstructor,\_,\_) \eqdef \uunit$$}%
  
  \smallskip
  \noindent Next, we define $\Inv$ to be $\false$ if any entry is $\mundef$ or the first entry is $\uunit$; otherwise, 
  \smallskip
  {\small\begin{align*}
  &\Inv((\exConstructor((\exc,\rc,x)),\ePending,e,\sPending,s)) ~\eqdef~ (\rc = \sPending + \agncount(s)) \land (\lnot \exc \implies \ePending = \uunit \land e = \uunit)\\
  &   \langindent\land (\exc \implies (\ePending = \exConstructor \lor e = \exConstructor) \land \lnot(\ePending = \exConstructor \land e = \exConstructor))
  \land (e = \exConstructor \implies s = \uunit)
      \land (\forall y, n .~ s = \agnConstructor(y, n) \implies x = y)\\
  &\text{(where $\agncount(\agnConstructor(x, n)) = n$ and $\agncount(\uunit) = 0$)}
  \end{align*}}%
  
  \smallskip
  These predicates can be stated in plain English:
the reference count $\rc$ is the total number of threads with the shared lock or in the process
of acquiring it; the $\exc$ field indicates whether any thread has
the exclusive lock or is in the process of acquiring it; an exclusive lock
cannot be taken at the same time as a shared lock; the value taken by a shared lock should match
the $\elemFields$'s $x$ value.

  Now, with the storage protocol established,
  we can embed these elements as propositions: we let
  $\Fields(\gamma,\exc,\rc,x) \eqdef \protOwn{\elemFields(\exc,\rc,x)}{\gamma}$
  and so on.
  We also let $\RwFamily(\gamma, F) \eqdef \maps(\gamma, F')$,
  where $F'(\exConstructor(x)) \eqdef F(x)$ and $F'(\uunit) \eqdef \true$.
  Now, in order to show our desired reader-writer lock rules (\autoref{fig:rwlock:logic}),
  it suffices to show the following updates (by \ruleref{SP-Update}):

  \smallskip
  {\small\begin{align*}
    \elemFields(\false, \rc, x) &\transitions \elemFields(\true, \rc, x) \cdot \elemExcPending
    \\
    \elemFields(\exc, \rc, x) &\transitions
      \elemFields(\exc, \rc+1, x) \cdot \elemSharedPending
    \\
    \elemFields(\false, \rc, x) \cdot \elemSharedPending
        &\transitions
          \elemFields(\false, \rc, x) \cdot \elemSharedGuard(x)
    \\
    \elemFields(\exc, \rc, x) \cdot \elemSharedGuard(y)
        &\transitions
            \elemFields(\exc, \rc - 1, x)
    \\
    \elemFields(\exc, \rc, x)  \cdot \elemSharedPending &\transitions
      \elemFields(\exc, \rc-1, x) 
  \end{align*}}%
  \smallskip
  
  \noindent As well as a withdraw (by \ruleref{SP-Withdraw}) and a deposit (by \ruleref{SP-Deposit}):
  \smallskip
  {\small\begin{align*}
    \elemFields(\exc, 0, x) \cdot \elemExcPending
        &\withdraws (\elemFields(\exc, 0, x) \cdot \elemExcGuard, \exConstructor(x))
        \\
        (\elemFields(\exc, \rc, y) \cdot \elemExcGuard, \exConstructor(x)) &\deposits \elemFields(\false, \rc, x) 
  \end{align*}}%
  \smallskip

  \noindent And finally, a guard (by \ruleref{SP-Guard}):
  {\small\begin{align*}
        \elemSharedGuard(x) &\guardsFO \exConstructor(x)
  \end{align*}}%
  Finally, we can prove all these just by expanding the definitions and using the logical invariants
  encoded in $\Inv$.

  Let us summarize exactly what the storage protocol gave us in this particular proof strategy.
  We wanted to construct some set of ghost resources with certain relationships, mostly updates ($\vs$), representing specific implementation details of the lock,
  with a single $\guards$ proposition that enables sharing. The storage protocol shows how to
  reduce those desired relationships to proof obligations ($\transitions$, $\withdraws$, $\deposits$, $\guardsFO$)
  about monoids that can be expressed in first-order logic.
  These obligations all encode properties that should map cleanly to an intuitive property of the system,
  e.g.,
  the $\withdraws$ proposition intuitively means ``from the intermediate pending state, if $\rc = 0$, then the stored resource can be withdrawn,''
  while
  $\elemSharedGuard(x) \guardsFO \exConstructor(x)$ intuitively means ``any reader agrees with the source-of-truth on what the shared value is.''
  These properties all rely on our definition of $\Sfunc$,
    a predicate that encodes which states of the system are well-formed.

\paragraph{Step 2: Verifying the implementation}
To verify the implementation (\autoref{fig:rwlock:impl}) against the spec (\autoref{fig:rwlock:spec}),
we first need to nail down a definition
for $\IsRwLock(\rw, \gamma, F)$.
Since $\gamma$ is meant to be the unique identifier
for the reader-writer lock, we can have it be the same as the ghost name $\gamma$
from the RwLock logic,
and likewise $F$, the family of propositions protected in the lock, be the same as $F$,
the family of propositions protected by the RwLock protocol.

The propositions representing the reader-writer lock should, as a whole, include
the proposition $\RwFamily(\gamma, F)$,
the permission to access the $\rw.\exc$ and $\rw.\rc$ memory cells, and the
$\Fields$ proposition which has the ghost data to match the contents of the memory cells.
{\small\begin{align*}
  &\IsRwLock(\rw, \gamma, F) \eqdef 
    \RwFamily(\gamma, F)
    * \exists \exc,\rc,x .~
    \Fields(\gamma, \exc, \rc, x)
    * (\rw.\exc \pointsto \exc)
    * (\rw.\rc \pointsto \rc)
\end{align*}}The proof for $\rwlocknew$ is then straightforward:
the implementation allocates the $\rw.\exc$ and $\rw.\rc$ memory,
and we can ghostily instantiate the RwLock protocol via
\ruleref{Rw-Init}.
Likewise, the proof for $\rwlockfree$ is straightforward,
since its precondition requires that the caller has exclusive access to $\IsRwLock$, so we can
destructure it and use the exclusive $\pointsto$ propositions in order to call $\langfree$.

The proofs for the other methods, which must operate over a \emph{shared}
$\IsRwLock$, are more interesting. Recall the proof outline for
$\acquireExc$:
\smallskip
{\small\begin{align*}
&\colorShared
[ \RwFamily(\gamma, F) * \exists \exc,\rc,x .~ \Fields(\gamma, \exc, \rc, x)
    * (\rw.\exc \pointsto \exc)
    * (\rw.\rc \pointsto \rc)
 ]\\
&
\colorShared \left\lfloor \color{black}
\begin{array}{ll}
  \{ ~ \}& \\
  \langdo&  \\
  \langindent \langlet ~ \textit{success} = \CAS(\rw.\exc, \false, \true)& \quad\quad(\ruleref{Rw-Exc-Begin}) \\
  \languntil ~ \textit{success}& \\
  \{ \ExcPending(\gamma) \}& \\
  \langdo& \\
  \langindent \langlet ~ r =~ \langbang \rw.\rc& \quad\quad(\ruleref{Rw-Exc-Acquire}) \\
  \languntil ~ r = 0& \\
  \{ \exists x.~ \ExcGuard(\gamma) * \later F(x) \} &
\end{array}\right.
\end{align*}}%
\smallskip

\noindent
The gist is that, in the first half, we apply \ruleref{Rw-Exc-Begin}
(in the case that the $\CAS$ succeeds)
to obtain $\ExcPending(\gamma)$, and in the second half, we apply \ruleref{Rw-Exc-Acquire}
to complete the acquisition and obtain the desired state $\ExcGuard(\gamma) * \later F(x)$.

Let us walk through the first half in detail. Since $\CAS$ is atomic,
we can apply \ruleref{Guard-Atomic-Inv} to ``open'' the shared proposition for the duration
of the atomic operation.  Thus, we need to show,
{\small\begin{align*}
&\{
  \RwFamily(\gamma, F) * \exists \exc,\rc,x .~ \Fields(\gamma, \exc, \rc, x)
    * (\rw.\exc \pointsto \exc)
    * (\rw.\rc \pointsto \rc)
\}\\
&\CAS(\rw.\exc, \false, \true)\\
&\{
  \textit{success}.~
  ((\textit{success}=\true * \ExcPending(\gamma)) \lor (\textit{success}=\false)) \:*\\
&  \phantom{xxxx}
  \RwFamily(\gamma, F) * \exists \exc,\rc,x .~ \Fields(\gamma, \exc, \rc, x)
    * (\rw.\exc \pointsto \exc)
    * (\rw.\rc \pointsto \rc)
\}
\end{align*}}%
If $\CAS$ succeeds, we have $\exc = \false$, so we apply $\ruleref{Rw-Exc-Begin}$.
This ensures we have $\ExcPending(\gamma)$ in the $\textit{success}=\true$ case.
Otherwise, we do nothing, and the program loops.
The second half of $\acquireExc$, where we atomically read $\rw.\rc$,
is the same, using \ruleref{Rw-Exc-Acquire}.

We can use a similar outline for $\acquireShared$:
{\small\begin{align*}
&\colorShared
[ \RwFamily(\gamma, F) * \exists \exc,\rc,x .~ \Fields(\gamma, \exc, \rc, x)
    * (\rw.\exc \pointsto \exc)
    * (\rw.\rc \pointsto \rc)
 ]\\
&
\colorShared \left\lfloor \color{black}
\begin{array}{ll}
  \{ ~ \}& \\
  \langdo & \\
  \langindent \{ ~ \}& \\
  \langindent \FetchAdd(\rw.\rc, 1);& \quad\quad (\ruleref{Rw-Shared-Begin}) \\
  \langindent \{ \SharedPending(\gamma) \}& \\
  \langindent \langlet ~ \exc =~ \langbang \rw.\exc ~ \langin & \quad\quad (\ruleref{Rw-Shared-Acquire})   \\
  \langindent \{ (\exc = \false * \exists x.~ \SharedGuard(\gamma, x)) \lor (\exc = \true * \SharedPending(\gamma)) \}& \\
  \langindent \langif ~ \exc ~ \langthen ~ \FetchAdd(\rw.\rc, -1);& \quad\quad (\ruleref{Rw-Shared-Retry}) \\
  \langindent \{ (\exc = \false * \exists x.~ \SharedGuard(\gamma, x)) \lor (\exc = \true) \}& \\
  \languntil ~ \exc = \false& \\
  \{ \exists x.~ \SharedGuard(\gamma, x) \}&
\end{array}\right.
\end{align*}}%
Finally, the proofs for $\acquireShared$, and $\releaseShared$
all follow similarly, using \ruleref{Rw-Exc-Release} and \ruleref{Rw-Shared-Release}
respectively.

\section{Hash Table Example: Composition with Shared State} \label{sec:ht}

In this section, we show how to use {\langname} to verify a larger application
built on top of the reader-writer lock developed in the previous section.
In particular, while the reader-writer lock gives us a mechanism to \emph{obtain}
and \emph{release} shared ghost state, here we will see how to \emph{make use} of shared state.
In particular, we show we can employ a fine-grained locking scheme, acquire multiple
shared locks, and compose state to perform nontrivial operations.
While the previous section was primarily an application of the storage protocol
(\autoref{sec:protocol}), this section will primarily be about applying the general-purpose
Leaf rules and ghost state (\autoref{sec:bg:pcm}-\autoref{sec:program-logic-heap}).

The example we consider is a concurrent \emph{linear probing hash table}~\cite{aocp-vol-3},
using a single lock per entry of the hash table.
We choose this example primarily because a single operation requires us to take multiple
locks, and in particular, for a query operation, we will be taking those locks in
shared, read-only mode. Thus, this example tests {\langname}'s ability as a logic for
manipulating shared resources.

\emph{Linear probing}, here, is a particular strategy a hash table uses to handle hash collisions.
Specifically, the hash table is arranged as an array with indices $0 \le i < L$, with some hash function $\hash : \Key \to [0, L)$.
To insert a key-value pair $(k, v)$, we attempt to insert it into the array at position $i = \hash(k)$.
If there is a different key already in the slot $i$, then we attempt to insert into $i+1$, and so on. Queries are similar: we scan starting at $i$ until we find the key or an empty slot.

In our implementation, we let $\htable$ be a record $\{\slots, \locks\}$
consisting of two arrays: one for the hash table slots, and one for the locks
protecting them. We then implement $\Query$ and $\Update$.

\begin{tabular}{cc}
{\begin{minipage}{0.35\textwidth}\small\begin{align*}
  &\Query(\htable,k) ~\eqdef~ \QueryIter(\htable,k,\hash(k))\\
  &\QueryIter ~\eqdef~ \langrec~\QueryIter(\htable,k,i).\\
      &\inde\langif~ i \ge L ~\langthen ~\langabort~ \langelse\\
      &\inde\inde\acquireShared(\htable.\locks[i]);\\
      &\inde\inde\langlet r = (\langmatch~ \langbang\htable.\slots[i] ~\langwith\\
      &\inde\inde\inde|~ \None \Rightarrow \None\\
      &\inde\inde\inde|~ \Some((k_i,v_i)) \Rightarrow\\
      &\inde\inde\inde\inde\langif~ k = k_i ~\langthen~\Some(v_i)\\
      &\inde\inde\inde\inde\langelse~ \QueryIter(\htable,k,i+1)\\
      &\inde\inde\langend) ~\langin\\
      &\inde\inde\releaseShared(\htable.\locks[i]);\\
      &\inde\inde r 
\end{align*}\end{minipage}}&{\begin{minipage}{0.35\textwidth}\small\begin{align*}
  &\Update(\htable,k) ~\eqdef~ \UpdateIter(\htable,k,\hash(k))\\
  &\UpdateIter ~\eqdef~ \langrec~\UpdateIter(\htable,k,i).\\
      &\inde\langif~ i \ge L ~\langthen ~\langabort~ \langelse\\
      &\inde\inde\acquireExc(\htable.\locks[i]);\\
      &\inde\inde\langmatch~ \langbang\htable.\slots[i] ~\langwith\\
      &\inde\inde\inde|~ \None \Rightarrow \htable.\slots[i] ~\langassign~ \Some((k, v))\\
      &\inde\inde\inde|~ \Some((k_i,v_i)) \Rightarrow\\
      &\inde\inde\inde\inde\langif~ k = k_i ~\langthen\\
      &\inde\inde\inde\inde\inde\htable.\slots[i] ~\langassign~ \Some((k, v))\\
      &\inde\inde\inde\inde\langelse~ \UpdateIter(\htable,k,i+1)\\
      &\inde\inde\langend\\
      &\inde\inde\releaseExc(\htable.\locks[i]);
\end{align*}\end{minipage}}\end{tabular}
\\
For simplicity, we do not handle hash table resizing (i.e., we assume a fixed $L$).
For the paper version only, the program aborts whenever a probe reaches the end of the table (index $L$).
This keeps the presentation manageable, though our full Coq version does gracefully handle the last case.

First, let us nail down the spec we want to prove.
There are a handful of options; here, we choose one
that manages the key-value mapping with propositions $\mapop(\gammaht, k, v)$,
(analogous to $\pointsto$ propositions).
As with the reader-writer lock, we also have a main proposition $\IsHT(\htable, \gammaht)$
to say that $\htable$ is a hash table, and since the hash table is concurrent,
the specifications for $\Query$ and $\Update$ require a shared
$\colorShared \IsHT(\htable, \gammaht)$.

\begin{center}\small
    \textbf{Hash Table Specification}

    \textbf{Propositions:}\quad
      $\IsHT(\htable, \gammaht)$\quad
      $\mapop(\gamma, k, v)$\quad
    \quad
    \text{(where $\htable: \Value,~~ \gammaht: \Name,~~ k: \Key,~~ v: \Value^?$)}

  \begin{mathpar}
    \{ \}~\htnew()~\{ \htable .~ \exists\gammaht, \IsHT(\htable, \gammaht) * 
        \MediumAsterisk{}_{k : \Key} ~ \mapop(\gamma, k, \None)
    \}
    \\

    \forall \htable,\gammaht .~
    \{ \IsHT(\htable, \gammaht) \}~\htfree(\htable)~\{ \}

    \forall \htable,\gammaht,k,v .~
    \hoareShared[ \IsHT(\htable, \gammaht) ]~
    \hoareShared[ \mapop(\gammaht, k, v) ]~
    \{ \}~\Query(\htable, k)~\{ v'.~ v = v' \}

    \forall \htable,\gammaht,v,v' .~
    \hoareShared[ \IsHT(\htable, \gammaht) ]~
    \{ \mapop(\gammaht, k, v) \}~\Update(\htable, k, v')~\{ \mapop(\gammaht, k, v') \}
  \end{mathpar}
\end{center}

In order to define the $\IsHT$ and $\mapop$ propositions and prove the specifications,
we once again start by defining a custom protocol of ghost state to represent the hash table's operation.

\paragraph{Step 1: Constructing the ghost resources} First, we create a custom ghost state
that lets us relate the contents of a hash table's slots to the key-value pairs
stored in the map, and which also encodes the appropriate kinds of state updates.
Here, we just use ``ordinary'' PCM ghost state (\autoref{sec:bg:pcm}), as we do not need to create any new $\guards$ relations.
Specifically, we can take the monoid
$(\Key \pfn \Ex(\Value^?)) \times (\mathbb{N} \pfn \Ex((\Key \times \Value)^?))$ and set,
{\small\begin{mathpar}
  \mapop(\gamma, k, v) \eqdef \ownGhost{\gamma}{([ k \mapsto \exConstructor(v)], [])}

  \slotop(\gamma, i, s) \eqdef \ownGhost{\gamma}{([], [ i \mapsto \exConstructor(s)])}
\end{mathpar}}%
For validity, $\V$, we encode invariants of the hash table: the key-value pairs
and slot entries are consistent, the slots obey hash-probing invariants, and so on.
(See \appref{app:ht} for full details.) We can then derive:
\begin{center}\small
    \textbf{Linear-Probing Hash Table Resource} 

    \textbf{Propositions:}\quad
        $\mapop(\gamma, k, v)$\quad
        $\slotop(\gamma, i, s)$
    \quad
    \text{(where
    $\gamma : \Name,~~
    k: \Key,~~
    v: \Value^?,~~
    i: \mathbb{N},~~
    s: (\Key \times \Value)^?$)}

    \begin{mathpar}
      \axiomhrefright{QueryFound}{QueryFound}{
        \mapop(\gamma,k,v) * \slotop(\gamma, j, \Some((k,v_j))) \entails v = \Some(v_j)
        \phantom{aaaa}
      }

      \inferhrefright{QueryNotFound}{QueryNotFound}{
        k \ne k_{\hash(k)},\ldots,k_{i-1}
      }{
          \mapop(\gamma,k,v) * \slotop(\gamma, i, \None)
          * (\MediumAsterisk{}_{\hash(k) \le j < i}~ \slotop(\gamma, j, \Some(k_j, v_j)))
          \entails v = \None
      }

      \axiomhrefright{UpdateExisting}{UpdateExisting}{
          \mapop(\gamma,k,v) * \slotop(\gamma, j, \Some((k,v_j)))
          \vs
          \mapop(\gamma,k,v') * \slotop(\gamma, j, \Some((k,v')))
        \phantom{aaaa}
      }

      \inferhrefright{UpdateInsert}{UpdateInsert}{
        k \ne k_{\hash(k)},\ldots,k_{j-1}
      }{
        \stacktwo{
          \mapop(\gamma,k,v) * \slotop(\gamma, i, \None)
          * (\MediumAsterisk{}_{\hash(k) \le j < i}~ \slotop(\gamma, j, \Some(k_j, v_j)))
        }{
          \vs
          \mapop(\gamma,k,v') * \slotop(\gamma, j, \Some((k,v')))
          * (\MediumAsterisk{}_{\hash(k) \le j < i}~ \slotop(\gamma, j, \Some(k_j, v_j)))
        }
      }
    \end{mathpar}
\end{center}

Now, what happens when we consider that the slot state $\slotop(\gamma, i, s)$
might be shared and read-only (via the reader-writer lock)?
Fortunately, the query-related rules can easily
be applied even if the $\slotop$ or $\mapop$ state is shared.
In particular, each query-related deduction concludes a pure proposition,
so we can apply \ruleref{Unguard-Pers} to obtain the predicate.
The update-related rules, meanwhile, can just be applied normally
using exclusively owned state
(though we could, in principle, apply \ruleref{Guard-Upd} to \ruleref{UpdateExisting}
if we needed to, since its $\MediumAsterisk$ term remains unchanged by the update).

There is only one essential capability that the Hash Table Resource lacks as written.
Specifically, we need to be able to \emph{compose} different pieces of state
from the Hash Table Resource.
For example, if we have shared propositions
$\colorShared \mapop(\gamma,k,v)$ and
$\colorShared \slotop(\gamma, j, \Some((k,v_j)))$, then
to apply \ruleref{QueryFound}, we actually need their \emph{composition},
$\colorShared \mapop(\gamma,k,v) * \slotop(\gamma, j, \Some((k,v_j)))$.
We can compose the shared state via \ruleref{Guard-And},
but this only gives us a $\land$ conjunction.
Therefore, we also need the following to complete the proofs for $\Query$ and $\Update$:
\begin{center} \small
    \textbf{Linear-Probing Hash Table Resource (Addendum)}
    \begin{mathpar}
        \mapop(\gamma,k,v) \land \slotop(\gamma,j,s)
        ~~\entails~~
        \mapop(\gamma,k,v) * \slotop(\gamma,j,s)

        \slotop(\gamma,b+1,s_{b+1})
        \land
        (\MediumAsterisk{}_{a \le j \le b}~ \slotop(\gamma,j,s_j))
        ~~\entails~~
            (\MediumAsterisk{}_{a \le j \le b + 1}~ \slotop(\gamma,j,s_j)))

        \mapop(\gamma,k,v)
        \land
        (\MediumAsterisk{}_{a \le j \le b}~ \slotop(\gamma,j,s_j))
        ~~\entails~~
        \mapop(\gamma,k,v) * 
            (\MediumAsterisk{}_{a \le j \le b}~ \slotop(\gamma,j,s_j))
    \end{mathpar}
\end{center}
Fortunately, in our PCM construction, these all follow from \ruleref{PCM-And}: in particular,
the hypothesis \ruleref{PCM-And} is satisfied for elements
$x$ and $y$ whenever the domains of the maps in $x$ and $y$ are disjoint.

\paragraph{Step 2: Verifying the implementation}

Once again, we need to establish a definition for $\IsHT(\gammaht, \htable)$, the
shareable proposition that makes something into a hash table.
Our hash table here is just made up of $L$ reader-writer locks:
{\small\begin{align*}
  \IsHT(\gammaht, \htable) \eqdef
      \MediumAsterisk{}_{0 \le i < L}~ \IsRwLock(\htable.\locks[i], \gamma_i, F_i)
\end{align*}}where we let $F_i = \lambda s.~ (\htable.\slots[i] \pointsto s) * \slotop(\gamma, i, s)$.
We also roll up all the relevant ghost names into the super-name
$\gammaht = (\gamma, \gamma_0, ...)$ to serve as a name for the hash table as a whole.

\begin{figure}
{\small$\begin{aligned}
  & \hoareShared[ \MediumAsterisk{}_{0 \le j < L}~ \IsRwLock(\htable.\locks[j]\colorShared, \gamma_j, F_j) ] ~ \hoareShared[ \mapop(k, v) ]\\
  & \hoareShared[ \MediumAsterisk{}_{\hash(k) \le j < i}~ \slotop(\gamma, j, \Some(k_j, v_j) * (k \ne k_j) ]\\
  &\colorShared \left\lfloor \color{black} \begin{array}{l}
      \inde\{ ~ \}\\
      \inde\langif~ i \ge L ~\langthen ~\langabort~\langelse\\
      \inde\inde\acquireShared(\htable.\locks[i]);\\
      \inde\inde\{ s.~ \SharedGuard(\gamma_i, s) \}\\
      \inde\inde\hoareShared[(\htable.\slots[i]\colorShared \pointsto s) * \slotop(\gamma, i, s)] \\
      \inde\inde\colorShared \left\lfloor \color{black} \begin{array}{l}
      \langlet r = (\langmatch~ \langbang\htable.\slots[i] ~\langwith\\
      \inde|~ \None \Rightarrow \None\\
      \inde|~ \Some((k_i,v_i)) \Rightarrow\\
      \inde\inde\langif~ k = k_i ~\langthen~\Some(v_i)\\
      \inde\inde\langelse~ \QueryIter(ht,k,i+1)\\
      \langend) ~\langin\\
        \end{array}\right.\\
      \inde\inde\{ \SharedGuard(\gamma_i, s) * (r = v) \}\\
      \inde\inde\releaseShared(\htable.\locks[i]);\\
      \inde\inde\{ r = v \}\\
      \inde\inde r 
  \end{array}\right.
\end{aligned}$}
  \vspace{-4mm}
  \caption{Proof outline of $\QueryIter$}\label{fig:proof:outline}
\end{figure}

The definition of $F_i$ effectively says that the $i$th lock protects both the
permission to access the $i$th slot of the hash table $\htable.\slots[i]$,
but also the slot state in the ghost protocol.

Crucially, composing the $\IsRwLock$ propositions with $*$ makes $\IsHT$ easy to work
with whether or not it is owned exclusively or shared.
In particular, if we have $\colorShared \IsHT(\gammaht, \htable)$ shared, as we expect of a concurrent hash
table, then we can use \ruleref{Guard-Split} to obtain a shared
$\colorShared \IsRwLock(\htable.\locks[i])$ so we can perform the usual lock operations
($\acquireShared$, and so on).

For the proof, we focus on $\Query$ here, as it makes the more
interesting use of shared state. The meat of $\Query$ is in
the recursive $\QueryIter$. Our proof of $\QueryIter$ is inductive, and it
takes in its precondition the ghost state for all the slots previously
accessed in the probe, in addition to the preconditions of $\Query$.
\autoref{fig:proof:outline} shows a proof outline.

Stepping through, we first call $\acquireShared$;
using our shared $\colorShared \IsRwLock(\htable.\locks[i]\colorShared, \gamma_i, F_i)$.
From this call, we obtain
$\SharedGuard(\gamma_i, s)$, for some $s$ which is fixed for the duration we hold the
lock.
By \ruleref{Rw-Shared-Guard} and the definition of $F_i$, we have
$\SharedGuard(\gamma, s) \guards[\gamma] \colorShared (\htable.\slots[i] \pointsto s) * \slotop(i, s)$
(eliminating the $\later$ by \ruleref{Later-Guard}). In the outline, we represent this shared state with our
$\colorShared [\ldots]$ notation.

Now, by \ruleref{LogicSec-Heap-Read-Shared} we can perform the $\langbang\htable.\slots[i]$ operation
to load the value of $s$ and case on it.
If the slot is empty ($\None$) then we apply \ruleref{QueryNotFound} to get our answer;
if the slot is full and the key matches, then we apply \ruleref{QueryFound}.
As discussed above, we have all we need to apply these deductions even when the state
on the left-hand side is shared.
The most interesting case is the recursive one: here, we append the newly obtained
$\colorShared \slotop(i, s)$ to obtain
$\colorShared \MediumAsterisk{}_{\hash(k) \le j \le i}~ \slotop(\gamma, j, \Some(k_j, v_j))$,
meeting the precondition for the recursive call.

\paragraph{The Client of the Hash Table}
There are many options available to the hash table's client.
We presume that the client wishes to share the hash table between threads,
and she has the freedom to do this as she wishes. For instance, she might
share it between a fixed number ($N$) of threads, using a fractional paradigm to give
out a fraction $1/N$ to each (\exampleref{ex:frac}).
Alternatively,
she could allocate it permanently and share it forever (\exampleref{ex:forever}).
Or she could put the hash table inside yet another reader-writer lock,
with multiple threads able to
concurrently access the hash table by taking a shared lock.
This last possibility could be a step to augment our design with resizing:
a client could take the lock exclusively to ``stop the world'' and rebuild the hash table.

\paragraph{Shared State with the Hash Table}
One might wonder if the client could further apply Leaf and use the hash table
to store propositions and manage shared access to them. For example, we might want
to say $\mapop(\gammaht, k, v) \guards F(k, v)$ for some proposition family $F$;
then any client with shared access to the key $k$ would also get shared access
to the resource $\colorShared F(k, v)$.
Indeed, we can modify our Hash Table Resource to allow this. Specifically,
we could reconstruct the resource via a storage protocol so we can prove $\guards$ propositions.
Effectively, the existing hash table monoid construction would become the protocol
monoid for this new storage protocol.

\section{More Advanced Storage Protocols}\label{sec:advanced}

The lock example in our paper is intentionally kept somewhat simple for the sake of exposition.
However, 
subsequent work has already used {\langname}'s storage protocols to 
verify far more sophisticated read-sharing mechanisms.
Specifically, IronSync~\cite{ironsync-tr} is a verification
framework that combines storage protocols with a handful of other techniques
(notably, using a substructural type system to manipulate ghost resources, including shared
ghost resources, rather than using CSL directly). 
Their framework embeds {\langname}'s monoidal storage protocol definitions as axioms
for manipulating their ghost resources.

They describe their experience using storage protocols to verify:
\begin{itemize}
  \item A multi-counter reader-writer lock with additional, domain-specific features.
    This is a component of an effort to verify a multi-threaded page cache;
    the lock is used to protect a \SI{4}{KiB} cache page, and the domain-specific features
    relate to reading and writing the cache page from disk.
    The lock not only allows read-sharing of memory resources for the \SI{4}{KiB} pages,
    but also of ghost resources related to their contents.
  \item A concurrent ring buffer with multiple producers and multiple consumers,
    where entries are alternately writeable and read-shared, as producer threads
    enqueue messages to be read (possibly simultaneously) by a number of consumer threads.
    This is a component of a state replication algorithm~\cite{nr-asplos17}, targeting
    non-uniform memory access (NUMA) architecture.
    Once again, this not only allows read-sharing of memory resources, but also ghost
    resources related to the operation log.
\end{itemize}
The second example, in particular, demonstrates that read-sharing protocols extend beyond 
reader-writer locks.
Furthermore, both examples (similar to our hash table)
demonstrate the use of read-shared custom ghost resources.

\section{Soundness} \label{sec:soundness}

Here, we sketch our construction of the {\langname} logic within the Iris separation logic;
for full details, consult the Coq development.
To get the most out of this section, it helps for the reader to be already familiar with Iris;
\citet{iris-from-the-ground-up} provide all necessary background.

\newcommand{\colorBase}[1]{{\color{blue}#1}}

As context, we review the components of Iris.
First, there is the \emph{Iris base logic}, a step-indexed logic of resources in the abstract,
with no primitive notion of a program or Hoare logic. The Iris base logic is proved
sound via a semantic model called \emph{the $\UPred$ model}.
Then, atop the base logic, Iris can do a variety of useful things, e.g., instantiate a program logic given some operational semantics.

To build {\langname}, we add a few minor deduction rules to the base logic, proved sound via the $\UPred$ model.
\colorBase{Our additions to the base logic are given in blue.}
We then define $\guards$, $\protOwn{p}{\gamma}$, and $\maps(\gamma, F)$,
and prove all of {\langname}'s deduction rules within the Iris logic.
The rest of the Iris framework, such as the machinery to instantiate a program logic and prove adequacy theorems, is unchanged.

\paragraph{Ghost state} In Iris, ghost state is constructed from a mathematical object called a \emph{CMRA}.
A PCM is a special case of a discrete CMRA, and the ghost state ($\ownGhost{\gamma}{x}$) in this paper is just the usual Iris ghost state.
We also \colorBase{add the \ruleref{PCM-And} rule to the base logic}, which follows straightforwardly from the definition of $\land$
over the $\UPred$ model and holds for any discrete CMRA.

\paragraph{Invariants}
Our definitions build on Iris's invariants, so we review those here.
Iris defines (within the base logic) a persistent proposition $\knowInv{\iota}{P}$ as
knowledge that an invariant $P$ is allocated at name $\iota$.
Iris then proves the following rules
so the user can allocate, open, and close invariants:
{\small\begin{align*}
  (\text{$\namesp$ infinite})&\entails \later P \vs[\mask] \exists \iota.~ (\iota \in \namesp) * \knowInv{\iota}{P}
    &\rulelabel{Inv-Alloc}\\
  (\iota\not\in\mask) * \knowInv{\iota}{P} &\entails \true \vs[\mask\cup\{\iota\}][\mask] (\later P)
    &\rulelabel{Inv-Open}\\
  (\iota\not\in\mask) * \knowInv{\iota}{P} &\entails (\later P) \vs[\mask][\mask\cup\{\iota\}] \true
    &\rulelabel{Inv-Close}
\end{align*}}%

\paragraph{The definition of $\guards$}
To define $\guards$, we first define a ``pre-guards'' operator, $\preguards$, for \emph{finite} sets $\zeta$,
and then take the closure under supersets to define the real $\guards$.
{\small\begin{align*}
    \AllocatedInvs(\zeta) &\eqdef
        \MediumAsterisk{}_{\iota \in \zeta} ~ \exists P:\iProp.~ \knowInv{\iota}{P}
    \\
    \OwnInvs(\zeta) &\eqdef
        \MediumAsterisk{}_{\iota \in \zeta} ~ \exists P:\iProp.~ \knowInv{\iota}{P} * P
    \\
    P \preguards[\zeta] Q &\eqdef 
        \AllocatedInvs(\zeta)
        \:*\\
    &\inde
    \forall R: \iProp .~ \always
    (
        P * (P \wand \OwnInvs(\zeta) * R)
        \wand \diamond~ (
        Q * (Q \wand \OwnInvs(\zeta) * R)
        )
    )
    \\
    P \guards[\mask] Q &\eqdef
        \exists \zeta.~ (\zeta \subseteq \mask) * (P \preguards[\zeta] Q)
\end{align*}}%
The $\preguards$ definition can be read roughly as, \emph{if we have the invariants at $\zeta$ and some other
state $R$, which can separate into $P$ and some other component, then it also can separate into $Q$}.
Note that this definition does not perform an update $(\upd)$ or even a mask change.
If we used $\upd$ we would not be able to show
\ruleref{Guard-And} because it would require us to perform two potentially contradictory
updates simultaneously.
It is also important that the definition preserves $\OwnInvs(\zeta)$, so that we can prove
\ruleref{Guard-Trans} without needing an additional disjointness condition.
Finally, taking the closure under supersets lets us prove \ruleref{Guard-Weaken-Mask}.

\paragraph{Basic $\guards$ proofs}
From this definition of $\guards$, many {\langname} rules are straightforward:
\ruleref{Guard-Refl},
\ruleref{Guard-Trans},
\ruleref{Guard-Split},
\ruleref{Guard-Weaken-Mask},
\ruleref{Guard-Pers}, and
\ruleref{Unguard-Pers}.
To prove \ruleref{Guard-Upd}, we perform a view shift from $\mask_1 \cup \mask_2$
to $\mask_1$, opening all the invariants~(\ruleref{Inv-Open}) in some $\zeta \subseteq \mask_2$.
This is primarily where we use the finiteness of $\zeta$ and the $\AllocatedInvs(\zeta)$
term.
\ruleref{Later-Guard} follows thanks to our use of $\diamond$ in the definition of $\guards$.
\ruleref{Later-Pers-Guard} then follows by pulling out the persistent part with
\ruleref{Unguard-Pers}, and then applying
\ruleref{Later-Guard}.

\paragraph{Point Propositions} Finally, we come to
\ruleref{Guard-Implies} and \ruleref{Guard-And}. The inherent difficulty here is that $A \entails P$ does not in general
imply $A$ can separate into $P * (P \wand A)$.
We therefore need to use the fact that $P$ is a point proposition.
The key is that point propositions are all of the form $\ownM{t}$ where $t$ is an element
of the \emph{global} CMRA instantiating the Iris model.
We can show:
{\small\colorBase{\begin{align*}
    A * (A \wand \ownM{t}) \entails
    \ownM{t} * (\ownM{t} \wand A)
  &\rulelabel{Split-Own}
\end{align*}}}%
This rule is somewhat peculiar: initially it looks like it must be wrong,
since applying $A$ and $A \wand P$ ought to give you just $P$, not $P$ plus something else.
On second glance, it makes some intuitive sense that you ought to be able to ``go back'' to $A$, since you had $A$ to begin with.
At any rate, it does at least hold for the $\ownM{t}$ propositions,
and it follows from the model definitions of $*$, $\wand$, and $\ownM{t}$.
\citet{reynolds-separation-logic-notes}
proved a similar theorem about what they called \emph{strictly exact assertions}; strictly exact assertions are not representable in Iris,
but $\ownM{t}$ propositions play a similar role.

To finish the proofs, we need something to account for our use of $\diamond$ in the definition of $\guards$.
{\small\colorBase{\begin{align*}
    A * (A \wand \diamond~ \ownM{t}) \entails
    (\diamond~ \ownM{t}) * (\ownM{t} \wand A)
  &\rulelabel{Split-Own-Except0}
\end{align*}}}%
From this, the proofs of \ruleref{Guard-Implies} and \ruleref{Guard-And} follow.

\paragraph{Storage Protocols}
Given an arbitrary storage protocol (\autoref{fig:protocol:definition}), we can define
a discrete CMRA $\Prot(P) \eqdef \init(P) ~|~ \uninit$ where $\uninit$ is the unit
and $\init(x) \cdot \init(y) = \init(x \cdot y)$.
We define validity on this algebra in terms of $\Inv$, specifically,
$\V(\uninit) = \true$ and $\V(\init(x)) = \exists y .~ \Inv(x \cdot y)$.
Then we construct ghost state via the authoritative-fragmentary construction,
$\Auth(\Prot(P))$. Let:
{\small\begin{align*}
    \maps(\gamma, F) &\eqdef \RespectsComposition(F)
        * \knowInv{\gamma}{\exists (x: P) .~ \ownGhost{\gamma}{\authfull~\init(x)} * \mathcal{C}(x) * F(\Sfunc(x))}
        * \ownGhost{\gamma}{\authfrag~\init(\uunit)}\\
    \protOwn{p}{\gamma} &\eqdef \ownGhost{\gamma}{\authfrag~\init(p)}
\end{align*}}In other words, the $\maps$ predicate gives us the invariant at location $\gamma$, which
contains the authoritative copy of some initialized protocol along with any state
currently stored in the protocol.

We can now proceed with the proofs of the laws in \autoref{fig:protocol:logic}.
To prove \ruleref{SP-Exchange}, we use our knowledge of the invariant from $\maps(\gamma, F)$ to open it,
obtaining the stored state $F(\Sfunc(x))$ and the authoritative knowledge of the protocol state,
$\ownGhost{\gamma}{\authfull~\init(x)}$. We perform the update and then close the invariant.
\ruleref{SP-Deposit},
\ruleref{SP-Withdraw}, and \ruleref{SP-Update} are just special cases
of \ruleref{SP-Exchange}.

For \ruleref{SP-Guard}, we need to prove a $\guards$ proposition.
Starting with the LHS of the $\preguards$ definition, we have
$\protOwn{p}{\gamma} * (\protOwn{p}{\gamma} \wand \OwnInvs(\zeta) * R)$.
This entails
$\ownGhost{\gamma}{\authfull~\init(x)} \land \ownGhost{\gamma}{\authfrag~\init(p)}$
for some $x$.
Using \ruleref{PCM-And}, we then obtain that $p \mle x$, apply the hypothesis
that $p \guardsFO s$, and pull $F(s)$ out of $F(\Sfunc(x))$.

\paragraph{Mechanization} The above definitions and proofs are formalized in Coq,
on top of the Iris library. Our Coq formalization also includes the instantiation of Leaf
on a heap-based language with atomic reads and writes (\autoref{sec:program-logic-heap}),
and the proof of the lock-based hash table (\autoref{sec:rwlock} and \autoref{sec:ht}).

\section{Related Work and Comparisons}\label{sec:related}

\paragraph{Shared, read-only ownership in separation logic}
Fractional permissions and counting permissions are existing mechanisms for temporarily shared read-only state that have appeared in a variety of contexts.
It is well-established that one can represent these via monoidal ghost state, but to our knowledge, the existing approaches do not provide a uniform way to reason about their interpretations as read-only state
the way {\langname}'s $\guards$ operator does. Below, we examine in depth what our case study would look like if we used traditional fractional resources.

Fractions have also been used in Iris's \emph{cancellable invariants},
i.e., invariants which are temporarily shared and then reclaimed, just as {\langname}'s guarded propositions are.
However, Iris's cancellable invariants do not support overlapping conjunction the way {\langname}'s $\guards$ does.

Fractions have also been used for a variety of more sophisticated applications, such as
RustBelt's lifetime logic~\cite{rustbelt} which can reason about Rust's shared borrows.
Again using fractions, \citet{rust-belt-relaxed-mem} show how to handle shared resource reclamation in relaxed memory settings.
We leave it as future work to see if {\langname}'s more general protocols can be adapted to those domains.

\citet{temporary-ro-permissions} introduce a ``temporary read-only modality'' for the sequential setting, allowing
a user to temporarily exchange a resource for a read-only resource, managed by
a lexical scope rule. In {\langname}, shared resources are not necessarily bound to a fixed
scope: guard propositions can be conditional and their scope dynamically determined.

Some prior work has also demonstrated more general permission logics capturing fractional and counting use cases~\cite{parkinsonPhD},
some with arbitrary composition structure~\cite{separation-algebras}.
However, these methods require \emph{all} state from the permission logic to 
be collected in order to regain exclusive access.
To our knowledge, they are not flexible enough to support protocols like our RwLock protocol,
which must represent nontrivial states even when no proposition is stored.
Meanwhile, fictional separation logic~\cite{fictional-separation-logic} uses a general extension mechanism similar to {\langname}'s relationship between protocol and storage monoids,
though it does not address general read-sharing mechanisms.

\paragraph{Concurrent separation logics with custom ghost state} 
Many CSL frameworks have introduced custom ghost state, i.e.,
mechanisms for users to define new resources,
such as the work on Concurrent Abstract Predicates (CAP)~\cite{ConcAbsPred,icap},
Fine-grained Concurrent Separation Logic (FCSL)~\cite{CommSTSFineGrained} with its concurroids (also based on PCMs),
and Iris with its CMRAs (see below).
To our knowledge, none of these frameworks have yet been used to provide a modular representation
of temporarily-shared custom resources that supports the composition of resources shared
via the representation.

\paragraph{Iris's ghost state formalism} 
{\langname} creates a custom ghost state mechanism, storage protocols, based on PCM ghost state.
Iris has generalized PCM ghost state in a different direction, creating an algebraic object
called a \emph{CMRA}~\cite{iris2}, which has two new features over PCMs.
The first is a built-in notion of \emph{persistent} state; in {\langname}, we de-emphasize persistent state because of our focus on temporarily-shared state.
However, it would be straightforward to incorporate persistence into the protocol monoid formalism.
The second is a \emph{step-indexed} notion of equality, which makes CMRAs suitable for \emph{higher-order} ghost state and many of the foundational elements of Iris.
In {\langname}, we wanted our storage protocols to be easily represented in first-order logic with a discrete notion of equality.
Factoring our map into two steps, $\Sfunc : P \to S$ and $F : S \to \iProp$ is what allows us to define a storage protocol without any step-indexing.
As such, one can understand storage protocols as a particular ghost state abstraction built on CMRA machinery.

\paragraph{Verified hash tables}
Hash tables have been verified before~\cite{fully-verified-container-library,pottier-ht-sequential-verification,aeneas-rust},
including a concurrent one done in Iris that uses mutual exclusion locks~\cite{iris-ht-thesis}.
Our concurrent hash table has some crucial differences that make it interesting:
\textbf{(i)}~we use reader-writer locks, and thus shared ownership for queries, and \textbf{(ii)}~ours requires a single operation (update or query) to take more than one lock.

\paragraph{Case study comparison}
It is worth comparing explicitly to how our reader-writer lock and hash table case study might be done if we used more traditional techniques. One of the most common such techniques in use to day is---as we have referenced several times in this paper---the technique of fractional permissions. If we were to build the hash table case study using fractional permissions (in Iris, or in any other framework supporting monoid ghost state and invariants),
it might look something like the following:

\begin{itemize}
    \item First, the resource being protected by the lock would need to have a built-in notion of being fractional. The reader-writer lock spec could be parameterized over a fractional proposition family, $F(x, q)$.
    The $\IsRwLock(\rw,\gamma,F)$ proposition would need to be fractionalized as well,
    which could be done using a technique called \emph{cancellable invariants}
    (invariants with associated fractional tokens, which allow the inner resources to be
    reclaimed).
    Ultimately, the lock's Hoare triples would look something like (to select a few):
{\small\begin{align*}
    \forall \rw,\gamma,F,q_0 .~ \hspace{0.2in}&\hspace{-0.2in}
    \{ \IsRwLock(\rw, \gamma, F, q_0) \}~\acquireShared(\rw)~\{ \IsRwLock(\rw, \gamma, F, q_0) * \exists x, q.~ \SharedGuard(\gamma, x, q) * F(x, q) \}
    \\
    \forall \rw,\gamma,F,x,q_0,q .~ &
    \{ \IsRwLock(\rw, \gamma, F, q_0) * \SharedGuard(\gamma, x, q) * F(x, q) \}~\releaseShared(\rw)~\{ \IsRwLock(\rw, \gamma, F, q_0) \} 
    \\
    \forall \rw, \gamma, F.~ &
    \{ \IsRwLock(\rw, \gamma, F, 1) \}~\rwlockfree(\rw)~\{ \} 
\end{align*}}%
Furthermore, the lock would need to guarantee that,
{\small\begin{align*}
\IsRwLock(\rw, \gamma, F, q_0) * \IsRwLock(\rw, \gamma, F, q_1) \bothentails \IsRwLock(\rw, \gamma, F, q_0 + q_1)
\end{align*}}%
while the client would have to promise that $F(x, q_0) * F(x, q_1) \bothentails F(x, q_0 + q_1)$.
Also observe that $\SharedGuard$ to track the fractional amounts that are ``lent out,'' so we can make sure the same amount is returned later.

    \item To verify the reader-writer lock,
        we would internally define an invariant that maintains some possibly-fractional
        amount of the resource, so that it has something to ``lend out'' whenever a client
        takes a read lock.
        Further, we would need to create ghost resources to
        define $\SharedGuard(\gamma, x, q)$, $\ExcGuard(\gamma)$, intermediate states, and so on.
        These resources would need to keep track
        of all the fractional amounts that are ``lent out,'' and make sure they sum
        to the correct amount.
        We would still need to reason about the intermediate states of the locking operations,
        but now the relationships are slightly harder to specify, because they interact
        with all of the fractional accounting.

    \item The client of the lock (the hash table) would need to make sure the resources it uses
        have a built-in fractional notion so they can interoperate with the lock.
        Thus the points-to operations would need a built-in notion of fractions
        $(\ell \pointstofrac{q} v)$ while the
        hash table's ``slot resources''
        $\slotop(\gamma, i, s)$ would be replaced by fractional resources
        $\slotop(\gamma, i, s, q)$.
        Now, all the updates and deductions would be expressed with fractions,
        and proving the $\transitions$ relations would involve reasoning about a composition
        operator $\cdot$ that adds rational numbers.
        For example, one has to reason like,
        ``Suppose I have a $q$ fraction of slot $j$, and a unit amount of slot
        $j+1$, and I replace slot $j+1$ with a unit amount of a new slot value...''
\end{itemize}

While this is all certainly possible, our perspective is that it involves a large number of
``bureaucratic'' details that do not directly relate to the programmer's primary intuition
of why the program is correct. By contrast, when doing this in the Leaf style, as we have seen:
\begin{itemize}
  \item The RwLock specification---that is, the ``interface'' between the two components---becomes cleaner. Neither $\IsRwLock$, $\SharedGuard$, nor $F$ need an additional rational
  number parameter~(\autoref{fig:rwlock:spec}). Instead, the relationships between these components and the sharing that takes place are all made clear
  through the $\guards$ and $\hoareShared[ \ldots ]~\{ \ldots \}~e~\{ \ldots \}$ notation.
  \item The RwLock is easier to verify because the storage protocol formulation helps us
      reduce the problem to a series of proof obligations regarding the evolution of
      the system.
  \item The Hash Table is easier to verify in Leaf because
      we can reason in a manner similar to how we would do it in an exclusive ownership setting,
      without the encoding having to ``bake in'' sharing-related details.
      For example, we would reason something like,
      ``Suppose I have slot $j$ and slot $j+1$, and I replace slot $j+1$...''
      and then rely on Leaf to apply this in the presence of shared slots.
      This sort of simplification would apply to any application that uses fine-grained reader-writer locks in a similar manner.
\end{itemize}

\section{Conclusion} \label{sec:conclusion}

We have introduced {\langname}, a concurrent separation logic with an approach to temporarily shared ownership based on our novel guarding operator, $\guards$.
We showed that {\langname} can help the user implement and verify sharing strategies,
that it allows modular specifications involving shared state that abstract away the sharing mechanism being used,
and that {\langname}'s composition capabilities allow it to handle fine-grained concurrency.

\section{Acknowledgments}
We thank Tej Chajed and the anonymous reviewers for helpful feedback.

Work at CMU was supported, in part, by an Amazon Research Award (Fall 2022
CFP), a gift from VMware, the Future Enterprise Security initiative at Carnegie
Mellon CyLab (FutureEnterprise@CyLab), and the NSF/VMware Partnership on
Software Defined Infrastructure as a Foundation for Clean-Slate Computing
Security (SDI-CSCS) program under Award No.\ CNS-1700521.


\newpage

\appendix

\section{Leaf (Technical Appendix)}

We include this appendix as a reference for additional technical content
that we did not have space for in the main paper.
We include:

\begin{itemize}
  \item Reference for Leaf rules and important Iris rules
          (as present in the main body of the paper).
  \item Concrete counterexamples to explain why simpler versions of some Leaf rules
        would not be sound.
  \item A basic ``counting permissions'' protocol.
  \item More details on the construction of the ``hash table logic.''
  \item A sketch of an expanded reader-writer lock with multiple reference counts.
  \item A sketch applying Leaf to heap semantics with non-atomic read and write operations.
\end{itemize}
Additional information can be found in the accompanying Coq formalization which can be found at \href{https://github.com/secure-foundations/leaf}{https://github.com/secure-foundations/leaf}, which includes:
\begin{itemize}
  \item Proofs for all the given {\langname} laws
  \item Program logic for an atomic heap-based language
  \item The forever, fractional, and counting protocols
  \item Full case studies from the paper, including implementation, ghost state, and proofs:
    \begin{itemize}
      \item The single-counter reader-writer lock
      \item The linear-probing hash table
    \end{itemize}
  \item Instantiation of the Iris adequacy theorem on the case studies
\end{itemize}


\newpage

\section{Leaf Deduction Rules}

\subsection{Iris Background: PCM Custom Ghost State}

Iris provides ghost state via an algebraic object called a \emph{CMRA} (sometimes \emph{camera}).
A PCM is a special case of a CMRA, and in particular, a PCM is a special case of
a \emph{discrete} CMRA (i.e., a CMRA whose equality does not depend on the step-index).
The version we present here is a simplified picture of Iris ghost state based on PCMs.
Note that \ruleref{XPCM-And} relies on the discreteness so that $\mle$ is well-defined.

Formally, a PCM is a set $M$ (the \emph{carrier} set) with a composition operator $\cdot : M \times M \to M$, and unit $\uunit : M$, and validity predicate $\V : M \to \bool$, where we
let,
{\small\begin{align*}
  a \mle b \eqdef \exists c.~ a \cdot c = b
\end{align*}}%
and,
{\small\begin{align*}
  \forall a,b.~& a \cdot b = b \cdot a\\
  \forall a,b,c.~& a \cdot (b \cdot c) = (a \cdot b) \cdot c\\
  \forall a.~& a \cdot \uunit = a\\
  \forall a,b.~& a \mle b \land \V(b) \implies \V(a)
\end{align*}}%

  \begin{center}\small
      \textbf{Rules for PCM ghost state}

      Instantiated for a given PCM $(M, \cdot, \V)$

      \textbf{Propositions:}\quad
          $\ownGhost{\gamma}{a}$
          \quad
      \text{(where
      $a : M,~~
      \gamma : \Name$)}

    \begin{mathpar}
      \axiomhref{PCM-Unit}{XPCM-Unit}{
      \true \entails \ownGhost{\gamma}{\uunit}
      }

      \axiomhref{PCM-Valid}{XPCM-Valid}{
      \ownGhost{\gamma}{a} \entails \V(a)
      }

      \axiomhref{PCM-Sep}{XPCM-Sep}{
      \ownGhost{\gamma}{a\cdot b}\bothentails\ownGhost{\gamma}{a} * \ownGhost{\gamma}{b}
      }

      \inferhref{PCM-Alloc}{XPCM-Alloc}{
        \V(a)
      }{
      \true \vs \exists\gamma.~ \ownGhost{\gamma}{a}
      }

      \inferhref{PCM-Update}{XPCM-Update}{
        a \transitions b
      }{
        \ownGhost{\gamma}{a} \vs \ownGhost{\gamma}{b}
      }

      \inferhref{PCM-And}{XPCM-And}{
          \forall t.~ ((x \mle t) \land (y \mle t) \land \V(t)) \implies (z \mle t)
      }{
          \ownGhost{\gamma}{x}
          \land
          \ownGhost{\gamma}{y}
          \entails
          \ownGhost{\gamma}{z}
      }
    \end{mathpar}
  \end{center}

\newpage

\subsection{Elementary $\guards$ deduction rules}

\begin{center}\small
  \textbf{Deduction rules for guarded resources}

  \textbf{Persistent Propositions:}\quad
    $P \guards[\mask] Q$
  \quad
  \text{(where $P, Q: \iProp,~~ \mask: \powset(\Name)$)}

  \begin{mathpar}
    \axiomhref{Guard-Refl}{XGuard-Refl}{
      P \guards[\mask] P
    }

    \axiomhref{Guard-Trans}{XGuard-Trans}{
      (P \guards[\mask] Q) *
      (Q \guards[\mask] R) \entails
      (P \guards[\mask] R)
    }

    \axiomhref{Guard-Split}{XGuard-Split}{
      P * Q \guards[\mask] P
    }

    \axiomhref{Guard-Weaken-Mask}{XGuard-Weaken-Mask}{
        (G \guards[\mask_1] P)
        \entails
        (G \guards[\mask_1 \cup \mask_2] P)
    }

    \inferhref{Guard-Pers}{XGuard-Pers}{
        \persistent{C}
    }{
        C \entails (\true \guards[\mask] C)
    }

    \inferhref{Unguard-Pers}{XUnguard-Pers}{
      A * B \entails C \quad
      \persistent{C}
    }{
      G * (G \guards[\mask] A) * B
      \vs[\mask]
      G * (G \guards[\mask] A) * B * C
    }

    \inferhref{Guard-Upd}{XGuard-Upd}{
        \mask_1 \cap \mask_2 = \emptyset
    }{
        (G \guards[\mask_1] P)
        *
        (P * A \vs[\mask_2] P * B)
        \entails
        (G * A \vs[\mask_1 \cup \mask_2] G * B)
    }

    \axiomhref{PointProp-Own}{XPointProp-Own}{
        \pointprop{\ownGhost{\gamma}{x}}
    }

    \inferhref{Guard-Implies}{XGuard-Implies}{
        A \entails P
        \quad
        \pointprop{P}
    }{
      (G \guards[\mask] A)
      \entails
      (G \guards[\mask] P)
    }

    \inferhref{Guard-And}{XGuard-And}{
        A \land B \entails P
        \quad
        \pointprop{P}
    }{
      (G \guards[\mask] A)
      *
      (G \guards[\mask] B)
      \entails
      (G \guards[\mask] P)
    }

    \inferhref{PointProp-Sep}{XPointProp-Sep}{
        \pointprop{P} \quad \pointprop{Q}
    }{
        \pointprop{P * Q}
    }
  \end{mathpar}
\end{center}

\newpage

\subsection{Storage Protocols}

\begin{figure}[H]
  \begin{subfigure}[t]{.48\textwidth}
  \begin{center}\small
  A \textbf{storage protocol} consists of:

  A \emph{storage monoid}, that is, a partial commutative monoid $(S, \cdot, \V)$, where, 
  \begin{align*}
      \forall a &.~ a \cdot \uunit = a\\
      \forall a, b &.~ a \cdot b = b \cdot a\\
      \forall a, b, c &.~ (a \cdot b) \cdot c = a \cdot (b \cdot c)\\
      &\phantom{.}~ \V(\uunit)\\
      \forall a, b &.~ a \mle b \land \V(b) \implies \V(a)
  \end{align*}

  A \emph{protocol monoid}, that is, a (total) commutative monoid $(P, \cdot)$, 
  with an arbitrary predicate $\Inv : P \to \bool$ and function $\Sfunc : P|_{\Inv} \to S$
  (i.e., the domain of $\Sfunc$ is restricted to the subset of $P$ where $\Inv$ holds)
  where,
  \begin{align*}
      \forall a &.~ a \cdot \uunit = a\\
      \forall a, b &.~ a \cdot b = b \cdot a\\
      \forall a, b, c &.~ (a \cdot b) \cdot c = a \cdot (b \cdot c)\\
      \forall a &.~ \Inv(a) \implies \V(\Sfunc(a))
  \end{align*}
  Note that $\Inv$ (unlike $\V$) is \emph{not} necessarily closed under $\mle$.
  ~\\
  \textbf{Additional notation} for storage protocols:\\
  ~\\
  For $p, p' : P$ and $s, s' : S$, define:
  \begin{align*}
      (p, s) \exchange (p', s') &\eqdef
          \forall q.~ \Inv(p \cdot q) \implies (\Inv(p' \cdot q)\\
            &
              \land \V(\Sfunc(p \cdot q) \cdot s)\\
            & \land \Sfunc(p \cdot q) \cdot s
                  = \Sfunc(p' \cdot q) \cdot s')\\
      p \withdraws (p', s') &\eqdef (p, \uunit) \exchange (p', s')\\
      (p, s) \deposits p'\phantom{(,s')} &\eqdef (p, s) \exchange (p', \uunit)\\
      p \transitions p'\phantom{(,s')} &\eqdef (p, \uunit) \exchange (p', \uunit)\\
      p \guardsFO s\phantom{(,p')} &\eqdef 
        \forall q.~ \Inv(p \cdot q) \implies s \mle \Sfunc(p \cdot q)
  \end{align*}
  \end{center}
  \caption{ \label{xfig:protocol:definition}
    Definition of a storage protocol and its derived relations.
  }
  \end{subfigure}
  \hfill
  \begin{subfigure}[t]{.48\textwidth}
  \begin{center}\small
    \textbf{Storage Protocol Logic}

    Instantiated for a given storage protocol $(S, \cdot, \V), (P, \cdot), \Inv, \Sfunc$

    \textbf{Propositions:}\quad
      $\protOwn{p}{\gamma}$

    \textbf{Persistent propositions:}\quad
      $\maps(\gamma, F)$

    \text{(where $\gamma: \Name,~~ F: S \to \iProp,~~ p: P$)}

    \begin{mathpar}
      \stacktwo{
      \RespectsComposition(F) ~~\eqdef~~
            \text{($F(\uunit) \bothentails \true$)\phantom{xxxxxx}}}
      {\text{and $\forall x,y.~ \V(x \cdot y) \implies (F(x \cdot y) \bothentails F(x) * F(y))$}}

      \inferhref{SP-Alloc}{XSP-Alloc}{
          \RespectsComposition(F)
          \quad
          \Inv(p)
          \quad
          \text{$\mathcal{N}$ infinite}
      }{
          F(\Sfunc(p))
          \vs
          \exists \gamma.~ \maps(\gamma, F) * \protOwn{p}{\gamma} * (\gamma \in \mathcal{N})
      }

      \inferhref{SP-Exchange}{XSP-Exchange}{
          (p,s) \exchange (p', s')
      }{
          \maps(\gamma, F)\entails
          (\later F(s)) * \protOwn{p}{\gamma}
          \vs[\gamma]
          (\later F(s')) * \protOwn{p'}{\gamma}
      }

      \inferhref{SP-Deposit}{XSP-Deposit}{
          (p,s) \deposits p'
      }{
          \maps(\gamma, F)\entails
          (\later F(s)) * \protOwn{p}{\gamma}
          \vs[\gamma]
          \protOwn{p'}{\gamma}
      }

      \inferhref{SP-Withdraw}{XSP-Withdraw}{
          p \withdraws (p', s')
      }{
          \maps(\gamma, F)\entails
          \protOwn{p}{\gamma}
          \vs[\gamma]
          (\later F(s')) * \protOwn{p'}{\gamma}
      }

      \inferhref{SP-Update}{XSP-Update}{
          p \transitions p'
      }{
          \maps(\gamma, F)\entails
          \protOwn{p}{\gamma}
          \vs[\gamma]
          \protOwn{p'}{\gamma}
      }

      \axiomhref{SP-PointProp}{XSP-PointProp}{
          \pointprop{\protOwn{p}{\gamma}}
      }

      \inferhref{SP-Guard}{XSP-Guard}{
          p \guardsFO s
      }{
          \maps(\gamma, F)\entails
          \protOwn{p}{\gamma}
          \guards[\gamma]
          (\later F(s))
      }

      \axiomhref{SP-Unit}{XSP-Unit}{
          \maps(\gamma, F) \entails \protOwn{\uunit}{\gamma}
      }

      \axiomhref{SP-Sep}{XSP-Sep}
      {
          \protOwn{p}{\gamma} *
          \protOwn{q}{\gamma}
          \bothentails
          \protOwn{p \cdot q}{\gamma}
      }

      \axiomhref{SP-Valid}{XSP-Valid}{
          \protOwn{p}{\gamma} \entails
              \exists q.~ \Inv(p \cdot q)
      }
    \end{mathpar}
  \end{center}
  \caption{ \label{xfig:protocol:logic}
      Deduction rules relating to a storage protocol and derived relations.
  }
  \end{subfigure}
\end{figure}

\newpage

\subsection{Remarks and Counterexamples} \label{app:counterexample}

We provide counterexamples to rules that one might initially expect, or hope, to be true.

\subsubsection{The restriction on \ruleref{XGuard-Implies}}

Suppose this rule held:

{\small\begin{mathpar}
    \inferhref{Wrong-Guard-Implies}{Wrong-Guard-Implies}{
        A \entails P
    }{
      (G \guards[\mask] A)
      \entails
      (G \guards[\mask] P)
    }
\end{mathpar}}%
We will derive a contradiction.
Take any propositions $P, Q$ such that
{\small\begin{mathpar}
\true \vs P * P

P * P \vs Q * Q

P * Q \vs \false
\end{mathpar}}%
Then we have,
{\small\begin{align*}
  P * (P \lor Q) &\vs P * P\\
                 &\vs Q * Q\\
                 &\vs Q * (P \lor Q)\\
\end{align*}}%
Now, $P \entails (P \lor Q)$ and so $P \guards (P \lor Q)$ and thus by \ruleref{XGuard-Upd},
{\small\begin{align*}
  P * P &\vs P * Q\\
\end{align*}}%
which lets us derive a contradiction,
{\small\begin{align*}
  \true
    &\vs P * P\\
    &\vs P * Q\\
    &\vs \false
\end{align*}}%

\subsubsection{The restriction on \ruleref{XGuard-And}}

Suppose this rule held:

{\small\begin{mathpar}
    \inferhref{Wrong-Guard-And}{Wrong-Guard-And}{
        A \land B \entails P
    }{
      (G \guards[\mask] A)
      *
      (G \guards[\mask] B)
      \entails
      (G \guards[\mask] P)
    }
\end{mathpar}}%

We will derive a contradiction.
Take any propositions $A, B, C$ such that,
{\small\begin{mathpar}
\true \vs A * B * C

A * A \entails \false

B * B \entails \false

C * C \entails \false

B \land C \entails B * C
\end{mathpar}}%
By \ruleref{XGuard-Split} we have,
{\small\begin{align*}
  (A \lor B) * (A \lor C)
  &\guards
  (A \lor B)
  \\
  (A \lor B) * (A \lor C)
  &\guards
  (A \lor C)
\end{align*}}%
So by \ruleref{Wrong-Guard-And}, we have,
{\small\begin{align*}
  (A \lor B) * (A \lor C) \guards
  (A \lor B) \land (A \lor C)
\end{align*}}%
Then we have,
{\small\begin{align*}
  A * ((A \lor B) \land (A \lor C))
    &\entails A * (B \land C)\\
    &\entails (B \land C) * A\\
    &\entails (B \land C) * ((A \lor B) \land (A \lor C))
\end{align*}}%
By \ruleref{XGuard-Upd}, we have
{\small\begin{align*}
  A * ((A \lor B) * (A \lor C))
    &\entails (B \land C) * ((A \lor B) * (A \lor C))
\end{align*}}
Finally,
{\small\begin{align*}
  \true
    &\vs A * B * C\\
    &\vs A * ((A \lor B) * (A \lor C))\\
    &\vs (B \land C) * ((A \lor B) * (A \lor C))\\
    &\vs (B * C) * ((A \lor B) * (A \lor C))\\
    &\vs \false
\end{align*}}

\subsubsection{The restriction on \ruleref{XGuard-Upd}}

Suppose this rule held:

{\small\begin{mathpar}
    \axiomhref{Wrong-Guard-Upd}{Wrong-Guard-Upd}{
        (G \guards[\mask_1] P)
        *
        (P * A \vs[\mask_2] P * B)
        \entails
        (G * A \vs[\mask_1 \cup \mask_2] G * B)
    }
\end{mathpar}}%
We will derive a contradiction.

Take fractional memory permissions.
We have $(\ell \pointstofrac{1/2} v) \guards[\NamespaceFrac] (\ell \pointsto v)$.
We also have,
{\small\begin{align*}
  (\ell \pointsto v) * (\ell \pointsto v) &\entails \false
\end{align*}}%
Therefore,
{\small\begin{align*}
  (\ell \pointsto v) * (\ell \pointsto v) &\vs \false * (\ell \pointsto v) * (\ell \pointsto v)
\end{align*}}%
We can apply \ruleref{Wrong-Guard-Upd} twice:
{\small\begin{align*}
  (\ell \pointstofrac{1/2} v) * (\ell \pointsto v) &\vs[\NamespaceFrac] \false * (\ell \pointstofrac{1/2} v) * (\ell \pointsto v)\\
  (\ell \pointstofrac{1/2} v) * (\ell \pointstofrac{1/2} v) &\vs[\NamespaceFrac] \false * (\ell \pointstofrac{1/2} v) * (\ell \pointstofrac{1/2} v)
\end{align*}}%
And so:
{\small\begin{align*}
  (\ell \pointstofrac{1} v) &\vs[\NamespaceFrac] \false
\end{align*}}%
from which we can derive a contradiction.

\subsubsection{The lack of any $*$ rule}

We might expect a rule like,
{\small\begin{mathpar}
\inferhref{Wrong-Guard-Sep-Disjoint}{Wrong-Guard-Sep-Disjoint}{
        \mask_1 \cap \mask_2 = \emptyset
    }{
        (G_1 \guards[\mask_1] P_1) * (G_1 \guards[\mask_2] P_2)
        \entails
        (G_1 * G_2) \guards[\mask_1 \cup \mask_2] (P_1 * P_2)
    }
\end{mathpar}}%
A special case would be,
\begin{mathpar}
    (G \guards[\mask] P) \entails (G * A \guards[\mask] P * A)
\end{mathpar}
In fact, not even the special case is sound.
To see why requires us to put an element of a storage protocol inside itself.
At a very high level: take any timeless proposition $P$ such that $P * P \entails \false$.
Insert $P$ into a storage protocol and obtain some $Q$ such that $Q \guards[\mask] P$.
Then insert $Q$ into the same protocol and obtain $R$ such that $R \guards[\mask] (P * Q)$.
Now, if the above rule held, we could get $(P * Q) \guards[\mask] (P * P)$.
Then by \ruleref{XGuard-Refl} we could get $R \guards[\mask] P * P$, which should be impossible.

Let us demonstrate in a little more detail that this argument really is possible.

It is not hard to construct a protocol, for any $X : \Set$ and $F : X \to \iProp$,
such that, upon initialization, we have the rules,
{\small\begin{align*}
    F(x) &\vs[\{\gamma\}] g(\gamma, x)\\
    g(\gamma, x) &\guards[\{\gamma\}]  \later F(x)\\
    g(\gamma, x_1) * g(\gamma, x_2) &\guards[\{\gamma\}]  (\later F(x_1)) * (\later F(x_2))
\end{align*}}%
where $g(\gamma, x)$ are duplicable propositions.
Let $X = \{\omega\} \cup \Name$ for some special value $\omega$.
Note that this is enough to define the storage protocol.

Therefore we can now define $F$. Now set:
{\small\begin{align*}
    F(\omega) \eqdef P
    \quad\quad
    F(\gamma) \eqdef g(\gamma, \omega)
\end{align*}}%

We can now deposit $F(\omega) = P$ to get $g(\gamma, \omega)$.
Call this new proposition $Q$.
We then have $Q \guards[\{\mask\}] P$.

Observe that $Q = g(\gamma, \omega) = F(\gamma)$. We can thus deposit $Q$ as well and get
$g(\gamma, \gamma)$. 

Finally, let $R = g(\gamma, \gamma) * g(\gamma, \omega)$.
We then have $R \guards[\{\gamma\}] P * Q$. This completes the outline above.

\newpage

\section{Derivation of a Counting Protocol} \label{app:count}

We show how to build a counting protocol, an analogue of Example 3.2 from the main paper,
which does a fractional protocol.

Specifically, let us suppose we have a single proposition, $Q$, we would like to manage.
We are not assuming that $Q * Q \entails \false$, which means that we might be able to
have multiple instances of $Q$ managed by the protocol at once, with multiple counters.

\newcommand{\rref}{\textlog{ref}}
\newcommand{\counter}{\textlog{counter}}

We want to construct a commutative monoid spanned by some elements $\rref$ and $\counter(n)$
modulo the identity $\rref \cdot \counter(n) = \counter(n-1)$. An easy way to do this is
to take the monoid,
{\small\begin{align*}
    P \eqdef \{ (r, c) : \mathbb{Z} \times \mathbb{N} ~|~ c = 0 \implies r \le 0 \}
\end{align*}}%
We define composition to simply be pairwise addition.
Let,
{\small\begin{align*}
  \rref &\eqdef (-1, 0)\\
  \counter(r) &\eqdef (r, 1)
\end{align*}}%
We want the counts from the $\rref$ objects to ``cancel out'' the counters' counts, so we set:
{\small\begin{align*}
  \Inv((r, c)) &\eqdef (r = 0)
\end{align*}}%
Finally, take the storage protocol to be $S \eqdef \mathbb{N}$, and the storage function to be
{\small\begin{align*}
  \Sfunc((r,c)) &\eqdef c
\end{align*}}%
One might notice in this set-up it is possible for counters to be negative, which seems a little
odd, but it does not matter since any negative counter would be canceled out by some other positive counter.

At any rate, we have,
{\begin{align*}
    \rref \cdot \counter(r) &= \counter(r-1)\\
    (\uunit, 1) &\exchange (\counter(0), 0)\\
    (\counter(0), 0) &\exchange (\uunit, 1)\\
    \rref &\guardsFO 1\\
\end{align*}}%
We could write these as:
{\begin{align*}
    (-1, 0) \cdot (r, 1) &= (r-1, 1)\\
    ((0, 0), 1) &\exchange ((0, 1), 0)\\
    ((0, 1), 0) &\exchange ((0, 0), 1)\\
    (-1, 0) &\guardsFO 1\\
\end{align*}}%
For the last one, it is crucial that we cannot write $(-1, 0) \cdot (1, 0) = (0, 0)$
because $(1, 0)$ is disallowed by the condition in the definition of $P$.
Observe that we had to include this condition in the definition of $P$ (rather than, say,
imposing it via $\Inv$) because the condition is not closed under $\mle$.

As such, if we have $(-1, 0) \cdot (r, c) = (0, c)$ then we must have $c > 0$, i.e.,
$c \ge 1$. Hence $(-1, 0) \guardsFO 1$.

\newpage

\section{Language Syntax and Program Logic}

In our case studies, we use a heap-based language with values and expressions given by:
\begin{align*}
v ~~:=~~&
    \true ~~|~~ \false
    ~~|~~ n
    ~~|~~ \langrec~ f(x).~ e
    ~~|~~ ()
    ~~|~~ (v_1,v_2)
    ~~|~~ \inl(v)
    ~~|~~ \inr(v)
    ~~|~~ \ell\\
e ~~:=~~&
    v
    ~~|~~ x
    ~~|~~ e_1 e_2
    ~~|~~ \langlet~x=e_1~\langin~e_2
    ~~|~~ e_1; e_2
    ~~|~~ (e_1,e_2)
    ~~|~~ \inl(e)
    ~~|~~ \inr(e)
    ~~|~~ \pi_i(e)
    ~~|~~ (\langmatch~e~\langwith \ldots)\\
    &~~|~~ (\langif~e_1~\langthen~e_2~\langelse~e_3)
    ~~|~~ \langfork~e
    ~~|~~ e_1 + e_2
    ~~|~~ e_1 = e_2\\
    ~~|~~ \langabort\\
    &~~|~~ \langref(e)
    ~~|~~ \langfree(e)
    ~~|~~ !e
    ~~|~~ e_1 ~\langassign~ e_2
    ~~|~~ \CAS(e_1,e_2,e_3)
    ~~|~~ \FetchAdd(e_1,e_2)
\end{align*}
Higher level constructs like records, arrays, or $\langdo ~\ldots~ \languntil$ can be considered
as syntactic sugar around a straightforward encoding.

We use standard operational semantics over a heap state
$\sigma : \Loc \fpfn \Value$.
In particular, we use the following head reduction steps for the heap operations:
\begin{align*}
  (\langref(v), \sigma) &\rightarrow (\ell, \sigma[\ell \mapsto v]) && \text{where $\ell \not\in \sigma$}\\
  (\langfree(\ell), \sigma) &\rightarrow ((), \sigma\backslash\{\ell\}) && \text{where $\ell \in \sigma$}\\
  (\langbang\ell, \sigma) &\rightarrow (v, \sigma) && \text{where $\ell \in \sigma$ and $\sigma(\ell) = v$}\\
  (\ell~\langassign~v, \sigma) &\rightarrow ((), \sigma[\ell \mapsto v]) && \text{where $\ell \in \sigma$}\\
  (\CAS(\ell, v_0, v_1), \sigma) &\rightarrow (\true, \sigma[\ell \mapsto v_1]) && \text{where $\ell \in \sigma$ and $\sigma(\ell) = v_0$}\\
  (\CAS(\ell, v_0, v_1), \sigma) &\rightarrow (\false, \sigma) && \text{where $\ell \in \sigma$ and $\sigma(\ell) \ne v_0$}\\
  (\FetchAdd(\ell, n), \sigma) &\rightarrow (m, \sigma[\ell \mapsto n + m]) && \text{where $\ell \in \sigma$ and $\sigma(\ell) = m$}\\
\end{align*}
Within Iris, using its standard method of defining the weakest-precondition and proving an adequacy theorem,
we can prove the following heap rules:
{\small\begin{mathpar}
\axiomhref{Heap-Ref}{Heap-Ref}{
  \inlinehoaresmall{ }~  \langref(v) ~\inlinehoaresmall{ \ell.~  \ell \pointsto v }
}

\axiomhref{Heap-Free}{Heap-Free}{
  \inlinehoaresmall{ \ell \pointsto v }~  \langfree(v) ~\inlinehoaresmall{ }
}

\axiomhref{Heap-Write}{Heap-Write}{
  \inlinehoaresmall{ \ell \pointsto v }~  \langbang\ell ~\inlinehoaresmall{ r.~  \ell \pointsto v * v = r }
}

\axiomhref{Heap-Read}{Heap-Read}{
  \inlinehoare{ \ell \pointsto v }~ \ell ~\langassign~ v' ~\inlinehoare{ \ell \pointsto v' }
}

\axiomhref{Heap-FetchAdd}{Heap-FetchAdd}{
  \inlinehoare{ \ell \pointsto n }~ \FetchAdd(\ell, m) ~\inlinehoare{ v.~ (v = m) * \ell \pointsto (n + m) }
}

\axiomhref{Heap-CAS-True}{Heap-CAS-True}{
  \inlinehoare{ \ell \pointsto v }~ \CAS(\ell, v, v') ~\inlinehoare{ r.~ (r = \true) * \ell \pointsto v' }
}

\axiomhref{Heap-CAS-False}{Heap-CAS-False}{
  \inlinehoare{ \ell \pointsto v * (v \ne v_1) }~ \CAS(\ell, v_1, v') ~\inlinehoare{ r.~ (r = \false) * \ell \pointsto v }
}
\end{mathpar}}%
We further define the shorthand:
{\small\begin{align*}
    \hoareShared[ X ][\mask]~\hoareShared[ \cdots ]~\{ P \}~e~\{ Q \}
     ~\eqdef~
    \forall G.~ \hoareShared[ \cdots ]~\{ P * G * (G \guards[\mask] X) \}~e~\{ Q * G \}
\end{align*}}
And can show:
{\small\begin{mathpar}
    \axiomhref{Heap-Read-Shared}{AppendixSec-Heap-Read-Shared}{
      \hoareShared[\ell \pointsto v]~
      \{  \} ~ \langbang\ell ~ \{ v'.~ v = v' \}
    }
\end{mathpar}}%

\newpage

\renewcommand{\acquireExc}{\textlog{multi\_lock\_exc}}
\renewcommand{\acquireShared}{\textlog{multi\_lock\_shared}}
\renewcommand{\releaseExc}{\textlog{multi\_unlock\_exc}}
\renewcommand{\releaseShared}{\textlog{multi\_unlock\_shared}}
\renewcommand{\rwlocknew}{\textlog{multi\_rwlock\_new}}
\renewcommand{\rwlockfree}{\textlog{multi\_rwlock\_free}}
\newcommand{\rcs}{\textit{rcs}}

\newcommand{\AgNVec}{\textsc{AgNVec}}

\section{Expanded Reader-Writer Lock: Multiple Reference Counters} \label{app:rwmulti}

The case study from the paper and the Coq formalization
uses a single reference counter.
Here, we sketch (on paper only) an expanded version for multiple reference counters.
The multiple reference counter version can be useful to reduce thread contention
in reader-heavy workloads.

Specifically, we consider the following implementation, for a fixed value $K$, the number
of counters. To acquire a read lock, the client supplies an integer $k \in [0, K)$ explaining
which counter they will use. To acquire an exclusive lock, the client has to check all counters.

\begin{align*}\small\begin{array}{l}
\rwlocknew() \eqdef\\
\langindent \langlet~ \rcs = \langfor ~ k ~ \langin ~ [0, K) ~ \langdo\\
\langindent\langindent \langref(0)\\
\langindent \langdone\\
\langindent \{ \exc: \langref(\false), \rcs: \rcs \}\\
  \\
\rwlockfree(\rw) \eqdef\\
  \langindent \langfree(\rw.\exc);\\
  \langindent \langfor ~ k ~ \langin ~ [0, K) ~ \langdo\\
  \langindent \langindent \langfree(\rw.\rcs[k])\\
  \langindent \langdone\\
  \langfree(\rw.\rcs)\\
  \\
\acquireExc(\rw) \eqdef\\
  \langindent \langdo \\
  \langindent \langindent \langlet ~ \textit{success} = \CAS(\rw.\exc, \false, \true)\\
  \langindent \languntil ~ \textit{success}\\
  \langindent \langfor ~ k ~ \langin ~ [0, K) ~ \langdo\\
  \langindent\langindent \langdo \\
  \langindent\langindent \langindent \langlet ~ r =~ \langbang \rw.\rcs\\
  \langindent\langindent \languntil ~ r = 0\\
  \langindent \langdone\\
  \\
\releaseExc(\rw) \eqdef\\
  \langindent \rw.\exc ~\langassign~ 0\\
  \\
\acquireShared(\rw, k) \eqdef\\
  \langindent \langdo \\
  \langindent \langindent \FetchAdd(\rw.\rcs[k], 1);\\
  \langindent \langindent \langlet ~ \exc =~ \langbang \rw.\exc ~ \langin \\
  \langindent \langindent \langif ~ \exc ~ \langthen ~ \FetchAdd(\rw.\rcs[k], -1);\\
  \langindent \languntil ~ \exc = \false\\
  \\
\releaseShared(\rw, k) \eqdef\\
  \langindent \FetchAdd(\rw.\rcs[k], -1)
\end{array}\end{align*}

\newpage

Our specification is mostly unchanged; we just need to account for $k$ in the shared locks.

\begin{center}\small
    \textbf{Multi-counter RwLock Specification}

    \textbf{Propositions:}\quad
      $\IsRwLock(\rw, \gamma, F)$\quad
      $\ExcGuard(\gamma)$\quad
      $\SharedGuard(\gamma, k, x)$\quad

    \text{(where $\rw: \Value,~~ \gamma: \Name,~~ X: \Set,~~ x: X,~~ F: X \to \iProp, k \in \mathbb{N}$ and $0 \le k < K$)}

  \small\begin{align*}
    \forall F,x.~ &&
    \{ F(x) \}~&\rwlocknew()&&~\{ \rw .~ \exists\gamma, \IsRwLock(\rw, \gamma, F) \} &
    \\
    \forall \rw, \gamma, F.~ &&
    \{ \IsRwLock(\rw, \gamma, F) \}~&\rwlockfree(\rw)&&~\{ \} &
    \\
    \forall \rw,\gamma,F .~ &
    \hoareShared[ \IsRwLock(\rw, \gamma, F) ]~&
    \{ \}~&\acquireExc(\rw)&&~\{ \ExcGuard(\gamma) * \exists x .~ \later F(x) \} &
    \\
    \forall \rw,\gamma,F,x .~ &
    \hoareShared[ \IsRwLock(\rw, \gamma, F) ]~&
    \{ \ExcGuard(\gamma) * F(x) \}~&\releaseExc(\rw)&&~\{ \} &
    \\
    \forall \rw,\gamma,F .~ &
    \hoareShared[ \IsRwLock(\rw, \gamma, F) ]~&
    \{ \}~&\acquireShared(\rw, k)&&~\{ \exists x.~ \SharedGuard(\gamma, k, x) \} &
    \\
    \forall \rw,\gamma,F,x .~ &
    \hoareShared[ \IsRwLock(\rw, \gamma, F) ]~&
    \{ \SharedGuard(\gamma, k, x) \}~&\releaseShared(\rw, k)&&~\{ \} &
    \\
    \forall \rw,\gamma,F,x.~ &
    \IsRwLock(\rw, \gamma, F) \entails (\SharedGuard(\gamma, k, x) \guards {\colorShared \later F(x)}) \hspace{-1in} &&&&&
  \end{align*}
\end{center}

To verify the implementation against the spec, it suffices to construct the following
reader-writer lock logic. Again, this is a variant on the one from the simpler lock.
The main difference is that $\SharedPending$ and $\SharedGuard$ both take an additional
parameter $k$ (the index of the counter being incremented)
and $\ExcPending$ takes an additional parameter $j$ (the number of counters that have
been checked so far).
We have one rule (\ruleref{XRw-Exc-Begin}) to enter the ``pending'' phase,
one rule (\ruleref{XRw-Exc-Progress}) to advance the counter $j$,
and one rule (\ruleref{XRw-Exc-Acquire}) to take the lock and perform the withdraw.

\begin{center}\small
\textbf{RwLock Storage Protocol Resource}

\textbf{Propositions:}\quad
    $\Fields(\gamma,\exc,\rcs,x)$
    \quad
    $\ExcPending(\gamma, j)$
    \quad
    $\ExcGuard(\gamma)$
    \quad
    $\SharedPending(\gamma, k)$
    \quad
    $\SharedGuard(\gamma, k, x)$

\textbf{Persistent Propositions:}\quad
    $\RwFamily(\gamma, F)$

  \text{(where
  $\gamma : \Name,~~
  \exc : \bool,~~
  \rcs : \mathbb{Z}^K,~~
  X : \Set,~~
  x : X,~~
  F : X \to \iProp, j, k: \mathbb{N}$ and $0 \le k < K$ and $0 \le j \le K$)}
\end{center}

{\small\begin{align*}
    F(x) &\vs
    \exists \gamma .~ \Fields(\gamma,\false,0,x)
        * \RwFamily(\gamma, F)
    &\rulelabelx{Rw-Init}\\
    \vspace{0.2em}
    \RwFamily(\gamma, F) \entails \hspace{0.5in} \\
    \Fields(\gamma, \false, \rcs, x) &\vs[\gamma] \Fields(\gamma, \true, \rcs, x) * \ExcPending(\gamma, 0)
    & \rulelabelx{Rw-Exc-Begin}\\
    \end{align*}
    \begin{align*}
    (j < K \land \rcs(j) = 0) *
    \Fields(\gamma, \exc, \rcs, x) * \ExcPending(\gamma, j)
        &\vs[\gamma]
        \Fields(\gamma, \exc, \rcs, x) * \ExcPending(\gamma, j + 1)
    & \rulelabelx{Rw-Exc-Progress}\\
    \end{align*}
    \begin{align*}
    \Fields(\gamma, \exc, \rcs, x) * \ExcPending(\gamma, K)
        &\vs[\gamma]
        \Fields(\gamma, \exc, \rcs, x) * \ExcGuard(\gamma) * \later F(x)
    & \rulelabelx{Rw-Exc-Acquire}\\
    \Fields(\gamma, \exc, \rcs, y) * \ExcGuard(\gamma) * \later F(x)
        &\vs[\gamma]
            \Fields(\gamma, \false, \rcs, x) 
    &\rulelabelx{Rw-Exc-Release}\\
    \Fields(\gamma, \exc, \rcs, x) &\vs[\gamma]
      \Fields(\gamma, \exc, \rcs[k \mapsto \rcs(k)+1], x) * \SharedPending(\gamma, k)
    &\rulelabelx{Rw-Shared-Begin}\\
    \Fields(\gamma, \false, \rcs, x) * \SharedPending(\gamma, k)
        &\vs[\gamma]
          \Fields(\gamma, \false, \rcs, x) * \SharedGuard(\gamma, k, x)
    &\rulelabelx{Rw-Shared-Acquire}\\
    \Fields(\gamma, \exc, \rcs, x) * \SharedGuard(\gamma, k, y)
        &\vs[\gamma]
            \Fields(\gamma, \exc, \rcs[k \mapsto \rcs(k) - 1], x)
    &\rulelabelx{Rw-Shared-Release}\\
    \Fields(\gamma, \exc, \rcs, x)  * \SharedPending(\gamma, k) &\vs[\gamma]
      \Fields(\gamma, \exc, \rcs[k \mapsto \rcs(k)-1], x) 
    &\rulelabelx{Rw-Shared-Retry}\\
    \SharedGuard(\gamma, k, x) &\guards[\gamma] \later F(x)
    &\rulelabelx{Rw-Shared-Guard}
    \end{align*}}

\newpage

Once again, we take the storage monoid $S$ to be $\Ex(X)$.
For the storage protocol construction, we first pick a monoid for each element:
\begin{itemize}
  \item For $\Fields$, we use $\Ex(\bool \times \mathbb{Z}^K \times X)$
  \item For $\ExcPending$, we use $\Ex([0, K])$
  \item For $\ExcGuard$, we use $\Ex(\UnitType)$
  \item For $\SharedPending$, we use $\mathbb{N}^K$
  \item For $\SharedGuard$, we use a variant on $\AgN(X)$ from the simpler version.
    Specifically, we define $\AgNVec(X)$ which uses a vector of counts instead of a single integer. Specifically, $\AgNVec(X)$ has elements,
      {\small\begin{align*}
          \uunit ~~|~~ \agnConstructor(x, n) ~~|~~ \mundef
      \end{align*}}%
  where $n \in \mathbb{N}^k$ and $n$ is nonzero.
  Once again, we let $\agnConstructor(x, n) \cdot \agnConstructor(x, m) = \agnConstructor(x, n+m)$,
  where $n+m$ is elementwise vector addition,
  and $\agnConstructor(x, n) \cdot \agnConstructor(y, m) = \mundef$ for $x \ne y$.
\end{itemize}

All in all, we take our protocol monoid $P$ to be the product,
{\small\begin{align*}
  \Ex(\bool \times \mathbb{Z}^K \times X)
  \times \Ex([0, K])
  \times \Ex(\UnitType)
  \times \mathbb{N}^K
  \times \AgNVec(X)
\end{align*}}%
Now we let,
{\small\begin{align*}\begin{array}{rlllll}
      \elemFields(\exc,\rcs,x) &\eqdef (\exConstructor((\exc,\rcs,x)),& \uunit,& \uunit,&  0,& \uunit)\\
      \elemExcPending(j) &\eqdef (\uunit,& \exConstructor(j),& \uunit,&  0,& \uunit)\\
      \elemExcGuard &\eqdef (\uunit,& \uunit,&  \exConstructor,& 0,& \uunit)\\
      \elemSharedPending(k) &\eqdef (\uunit,& \uunit,&  \uunit,&  [k \mapsto 1],& \uunit)\\
      \elemSharedGuard(k) &\eqdef (\uunit,& \uunit,&  \uunit,&  0,& \agnConstructor(x, [k \mapsto 1]))
  \end{array}\end{align*}}%
  Let
  $\Fields(\gamma,\exc,\rcs,x) \eqdef \protOwn{\elemFields(\exc,\rcs,x)}{\gamma}$
  and so on,
  and also let $\RwFamily(\gamma, F) \eqdef \maps(\gamma, F)$.

  We define $\Sfunc$:
    {\small\begin{align*}
      \Sfunc(\exConstructor((\exc,\rcs,x)),\_,\uunit,\_,\_) &\eqdef \exConstructor(x)\\
      \Sfunc(\exConstructor((\exc,\rcs,x)),\_,\exConstructor,\_,\_) &\eqdef \uunit
  \end{align*}}

  We define $\Inv$ to be $\false$ if any entry is $\mundef$ or the first entry is $\uunit$; otherwise, 
  {\small\begin{align*}
  &\Inv((\exConstructor((\exc,\rcs,x)),\ePending,e,\sPending,s)) ~\eqdef~ (\forall k.~ \rcs(k) = \sPending(k) + \agncount(k, s)) \land (\lnot \exc \implies \ePending = \uunit \land e = \uunit)\\
  &   \langindent\land (\exc \implies (\ePending \ne \uunit \lor e \ne \uunit) \land \lnot(\ePending \ne \uunit  \land e \ne \uunit))
  \land (e \implies s = \uunit)
      \land (\forall y, n .~ s = \agnConstructor(y, n) \implies x = y)\\
  &   \langindent\land (\forall k,k' .~ \ePending = \exConstructor(k') \land k < k' \implies \agncount(k, s) = 0)\\
  &\text{(where $\agncount(k, \agnConstructor(x, n)) = n(k)$ and $\agncount(k, \uunit) = 0$)}
  \end{align*}}
Now, we can show:
\begin{itemize}
  \item \ruleref{XRw-Exc-Acquire} by \ruleref{XSP-Withdraw}
  \item \ruleref{XRw-Exc-Release} by \ruleref{XSP-Deposit}
  \item \ruleref{XRw-Shared-Guard} by \ruleref{XSP-Guard}
  \item The rest via \ruleref{XSP-Update}
\end{itemize}

\newpage

%
%
%
%

\section{Hash Table Custom Ghost State Construction} \label{app:ht}

Fix an integer $L > 0$ and a hash function $\hash : \Key \to \mathbb{N}$.
Consider the monoid
\[
  (\Key \pfn \Ex(\Value^?)) \times (\mathbb{N} \pfn \Ex((\Key \times \Value)^?)).
\]
Define validity by:
{\small\begin{align*}
  \KeysDistinct((\sigma,\mu)) &~\eqdef~ \forall i_1,k_1,v_1,i_2,v_2 .~
      \sigma(i_1) = \exConstructor((k_1, v_1)) \land \sigma(i_2) = \exConstructor((k_2, v_2)) \land i_1 \ne i_2 \implies
          k_1 \ne k_2\\
  \MapConsistent((\sigma,\mu)) &~\eqdef~ \forall k,v .~
      \mu(k) = \exConstructor(\Some(v)) \implies \exists i.~ \sigma(i) = \exConstructor(\Some((k,v)))\\
  \SlotsConsistent((\sigma,\mu)) &~\eqdef~ \forall i,k,v .~
      \sigma(i) = \exConstructor(\Some((k, v))) \implies k \in \mu \land \mu(k) \ne \exConstructor(\None)\\
  \Contiguous((\sigma,\mu)) &~\eqdef~ \forall i,k,v .~
      \mu(i) = \exConstructor(\Some((k,v))) \implies \hash(k) \le i\\
      &\langindent\langindent
        \land
        (\forall j.~ \hash(k) \le j \le i \implies j \in \mu \land \mu(j) \ne \exConstructor(\None))\\
  \\
  P((\sigma,\mu)) &~\eqdef
    (\forall i .~ i \in \sigma \implies \exists s.~ \sigma(i) = \exConstructor(s))\\
    &\langindent\langindent\land (\forall k .~ k \in \mu \implies \exists v.~ \mu(k) = \exConstructor(v))\\
    &\langindent\langindent
        \land\KeysDistinct((\sigma,\mu))\\
    &\langindent\langindent
        \land \MapConsistent((\sigma,\mu))\\
    &\langindent\langindent
      \land \SlotsConsistent((\sigma,\mu))\\
    &\langindent\langindent
      \land \Contiguous((\sigma,\mu))\\
  \V(z) &~\eqdef~ \exists z'. ~ z \mle z' \land P(z')\\
\end{align*}}%
These invariants can be interpreted as follows:
\begin{itemize}
  \item $\KeysDistinct$ says all slots in the hash table have distinct keys.
  \item $\MapConsistent$ says that for each key-value pair $(k, v)$ in the map, there is a table entry containing $(k, v)$.
  \item $\SlotsConsistent$ says that for each entry with a key-value pair $(k, v)$, that pair is in the map.
  \item $\Contiguous$ says that each key in the table, there is a contiguous range of nonempty entries from the key's hash to that entry.
\end{itemize}
Now set,
{\small\begin{mathpar}
  \mapop(\gamma, k, v) \eqdef ([ k \mapsto \exConstructor(v)], [])

  \slotop(\gamma, i, s) \eqdef ([], [ i \mapsto \exConstructor(s)])
\end{mathpar}}%
Now we prove the following:
\begin{center}\small
    \textbf{Linear-Probing Hash Table Resource} 

    \textbf{Propositions:}\quad
        $\mapop(\gamma, k, v)$\quad
        $\slotop(\gamma, i, s)$
    \quad
    \text{(where
    $\gamma : \Name,~~
    k: \Key,~~
    v: \Value^?,~~
    i: \mathbb{N},~~
    s: (\Key \times \Value)^?$)}

    \begin{mathpar}
      \axiomhrefright{QueryFound}{XQueryFound}{
        \mapop(\gamma,k,v) * \slotop(\gamma, j, \Some((k,v_j))) \entails v = \Some(v_j)
        \phantom{aaaa}
      }

      \inferhrefright{QueryNotFound}{XQueryNotFound}{
        k \ne k_{\hash(k)},\ldots,k_{i-1}
      }{
          \mapop(\gamma,k,v) * \slotop(\gamma, i, \None)
          * (\MediumAsterisk{}_{\hash(k) \le j < i}~ \slotop(\gamma, j, \Some(k_j, v_j)))
          \entails v = \None
      }

      \axiomhrefright{UpdateExisting}{XUpdateExisting}{
          \mapop(\gamma,k,v) * \slotop(\gamma, j, \Some((k,v_j)))
          \vs
          \mapop(\gamma,k,v') * \slotop(\gamma, j, \Some((k,v')))
        \phantom{aaaa}
      }

      \inferhrefright{UpdateInsert}{XUpdateInsert}{
        k \ne k_{\hash(k)},\ldots,k_{j-1}
      }{
        \stacktwo{
          \mapop(\gamma,k,v) * \slotop(\gamma, i, \None)
          * (\MediumAsterisk{}_{\hash(k) \le j < i}~ \slotop(\gamma, j, \Some(k_j, v_j)))
        }{
          \vs
          \mapop(\gamma,k,v') * \slotop(\gamma, j, \Some((k,v')))
          * (\MediumAsterisk{}_{\hash(k) \le j < i}~ \slotop(\gamma, j, \Some(k_j, v_j)))
        }
      }

      \mapop(\gamma,k,v) \land \slotop(\gamma,j,s)
        ~~\entails~~
        \mapop(\gamma,k,v) * \slotop(\gamma,j,s)

        \slotop(\gamma,b+1,s_{b+1})
        \land
        (\MediumAsterisk{}_{a \le j \le b}~ \slotop(\gamma,j,s_j))
        ~~\entails~~
            (\MediumAsterisk{}_{a \le j \le b + 1}~ \slotop(\gamma,j,s_j)))

        \mapop(\gamma,k,v)
        \land
        (\MediumAsterisk{}_{a \le j \le b}~ \slotop(\gamma,j,s_j))
        ~~\entails~~
        \mapop(\gamma,k,v) * 
            (\MediumAsterisk{}_{a \le j \le b}~ \slotop(\gamma,j,s_j))
    \end{mathpar}
\end{center}
Specifically,
\begin{itemize}
  \item We can prove \ruleref{XQueryFound} and \ruleref{XQueryNotFound} by \ruleref{XPCM-Valid}.
  \item We can prove \ruleref{XUpdateExisting} and \ruleref{XUpdateInsert} by \ruleref{XPCM-Update}.
  \item We can prove the last three rules by \ruleref{XPCM-And}.
\end{itemize}

\newpage

\newcommand{\orderingNAInter}{\textlog{na'}}
\newcommand{\orderingNA}{\textlog{na}}
\newcommand{\orderingSC}{\textlog{sc}}

\newcommand{\langbangNAInter}{\langitem{!_{\orderingNAInter}}}
\newcommand{\langbangNA}{\langitem{!_{\orderingNA}}}
\newcommand{\langbangSC}{\langitem{!_{\orderingSC}}}

\newcommand{\langassignNAInter}{{\color{blue}\leftarrow_{\orderingNAInter}}}
\newcommand{\langassignNA}{{\color{blue}\leftarrow_{\orderingNA}}}
\newcommand{\langassignSC}{{\color{blue}\leftarrow_{\orderingSC}}}

\newcommand{\ReadWriteState}{\textsf{ReadWrite}}
\newcommand{\RWState}{\textsf{ReadWrite'}}
\newcommand{\stateReading}[1]{\textsf{reading}(#1)}
\newcommand{\stateWriting}{\textsf{writing}}

\newcommand{\stateW}{\textsf{writing}}
\newcommand{\stateR}{\textsf{reading}}

\newcommand{\gammaHeap}{\ensuremath{\gamma_{\text{heap}}}}
\newcommand{\gammaAux}{\ensuremath{\gamma_{\text{reads}}}}

\newcommand{\countsAgree}{\textit{countsAgree}}

\newcommand{\pointstoInter}[2]{#1\xhookrightarrow{\smash{\raisebox{-.3ex}{\ensuremath{\scriptstyle\kern-0.25ex\textlog{w}\kern-0.1ex}}}} #2}

\newcommand{\maskHeap}{\mask_{\text{heap}}}

\section{Language Extension for Non-Atomic Memory} \label{app:nonatomic}

In this section, we sketch how {\langname} can be applied to a language that includes
non-atomic memory.

First, we establish the language and its semantics. We focus here only on heap-related
aspects of the language.
Each read or write operation is annotated with a label, $\orderingNA$ (non-atomic) or $\orderingSC$ (sequentially-consistent atomic).
As the names suggest, the atomic read ($\langbangSC$) and atomic write ($\langassignSC$)
are each executed in a single atomic operation.
Meanwhile, the non-atomic operations 
($\langbangNA$ and $\langassignNA$)
each take two steps.
\begin{align*}
e ~~:=~~&
    ~~|~~ e_1 ~\langassignSC~ e_2
    ~~|~~ \langbangSC e
    \\
    &
    ~~|~~ e_1 ~\langassignNA~ e_2
    ~~|~~ \langbangNA e
    \\
    &
    ~~|~~ e_1 ~\langassignNAInter~ e_2
    ~~|~~ \langbangNAInter e
    \\
    & \cdots
\end{align*}
The $\orderingNAInter$ operations are ``artificial'' language elements, used as
intermediate states for the operational semantics.

The heap state is given by $\sigma : \Loc \fpfn (\Value \times \ReadWriteState)$,
where the set $\ReadWriteState$ is given by the elements
\[
    \stateReading{n}
    ~~|~~
    \stateWriting
\]
where $n : \mathbb{N}$.
This state is used to track the in-progress non-atomic operations.
The state $\stateWriting$ means that a (single) write is in-progress,
while $\stateReading{n}$ means that $n$ reads are in-progress.
Of course, this means that $\stateReading{0}$ means that no reads or writes are in-progress.

By tracking this state, our heap semantics are able to reason about \emph{data races}.
These occur when either \textbf{(i)} a non-atomic read overlaps with a (atomic or non-atomic)
write or \textbf{(ii)} a non-atomic write overlaps with any other operation.
Any such data-race will result in a ``stuck state.''

The execution semantics are given as follows:
\begin{align*}
  (\langbangSC\ell, \sigma[ \ell \mapsto (v, \stateReading{n})]) &\rightarrow (v, \sigma[ \ell \mapsto (v, \stateReading{n}) ]) \\
  & \\
  (\ell~\langassignSC~v', \sigma[ \ell \mapsto (v, \stateReading{0})]) &\rightarrow ((), \sigma[\ell \mapsto (v', \stateReading{0})]) \\
  & \\
  (\langbangNA\ell, \sigma[ \ell \mapsto (v, \stateReading{n})]) &\rightarrow (\langbangNAInter\ell, \sigma[ \ell \mapsto (v, \stateReading{n + 1})])  \\
  (\langbangNAInter\ell, \sigma[ \ell \mapsto (v, \stateReading{n + 1})]) &\rightarrow (v, \sigma[ \ell \mapsto (v, \stateReading{n})])  \\
  & \\
  (\ell~\langassignNA~v', \sigma[ \ell \mapsto (v, \stateReading{0})]) &\rightarrow (\ell~\langassignNAInter~v', \sigma[ \ell \mapsto (v, \stateWriting)]) \\
  (\ell~\langassignNAInter~v', \sigma[ \ell \mapsto (v, \stateWriting)])
  &\rightarrow
      ((), \sigma[ \ell \mapsto (v', \stateReading{0})])
\end{align*}
If we cannot execute the appropriate rule because the $\ReadWriteState$ state is wrong, we result in a stuck state.

We want to prove the usual Hoare triples for both $\orderingSC$ and $\orderingNA$ operations.

{\small\begin{mathpar}
\axiomhref{Heap-Write-SC}{Heap-Write-SC}{
  \inlinehoaresmall{ \ell \pointsto v }~  \langbangSC\ell ~\inlinehoaresmall{ r.~  \ell \pointsto v * v = r }_{\maskHeap}
}

\axiomhref{Heap-Read-Shared-SC}{Heap-Read-Shared-SC}{
  \hoareShared[\ell \pointsto v][\maskHeap]~
  \{  \} ~ \langbangSC\ell ~ \{ v'.~ v = v' \}_{\maskHeap}
}
\end{mathpar}}%

{\small\begin{mathpar}
\axiomhref{Heap-Write-NA}{Heap-Write-NA}{
  \inlinehoaresmall{ \ell \pointsto v }~  \langbangNA\ell ~\inlinehoaresmall{ r.~  \ell \pointsto v * v = r }_{\maskHeap}
}

\axiomhref{Heap-Read-Shared-NA}{Heap-Read-Shared-NA}{
  \hoareShared[\ell \pointsto v][\maskHeap]~
  \{  \} ~ \langbangNA\ell ~ \{ v'.~ v = v' \}_{\maskHeap}
}
\end{mathpar}}%

From the perspective of the program logic, the only difference between $\orderingSC$
and $\orderingNA$ operations is that the $\orderingNA$ operations not physically atomic.

Regardless, we are here to prove the Hoare triples.
Keep in mind that proving that the triples requires us to prove that the operations
do not get stuck when we have the appropriate resources.
The key challenge here is that
the ``read'' operations are actually effectful, temporarily updating the $\ReadWriteState$
at the given location. Therefore, it seems difficult to make use of 
a shared $\colorShared \ell \pointsto v$. The trick is to define $\HeapInterp(\sigma)$
and $\ell \pointsto v$ the right way.

First, define the set $\RWState$ with the elements
\[
    \stateR
    ~~|~~
    \stateW
\]
Define $t : \ReadWriteState \to \RWState$ by,
\begin{align*}
    t(\stateReading{n}) ~~\eqdef~~ \stateR\\
    t(\stateWriting) ~~\eqdef~~ \stateW
\end{align*}
Essentially, this tracks the writing versus non-writing states, while ``forgetting''
the heap reference count.
Now define $T : (\Loc \fpfn (\Value \times \ReadWriteState)) \to 
    (\Loc \fpfn (\Value \times \RWState))$
by mapping $t$ over each $\ReadWriteState$.

Now we use the authoritative-fragmentary construction,
\[
\Auth( \Loc \fpfn (\Value \times \RWState) ) \text{~for ghost state at~} \gammaHeap,
\]
and the (higher order) construction,
\[
\Auth( (\Loc \times \mathbb{N}) \fpfn \latert(\iProp) ) \text{~for ghost state at~} \gammaAux.
\]
Now define,
{\small\begin{align*}
    \ell \pointsto v &~~\eqdef~~
        \ownGhost{\gammaHeap}{\authfrag~[ \ell \mapsto (v, \stateR) ]}
        \\
    \pointstoInter{\ell}{v} &~~\eqdef~~
        \ownGhost{\gammaHeap}{\authfrag~[ \ell \mapsto (v, \stateW) ]}
        \\
    \HeapInterp(\sigma) &~~\eqdef~~
        \exists (m : (\Loc \times \mathbb{N}) \fpfn \iProp) .~\\
        &\ownGhost{\gammaHeap}{\authfull~T(\sigma)}
        *
            \ownGhost{\gammaAux}{\authfull~\latertinj(m)}
            *
            \countsAgree(\sigma, m)
            *
            \MediumAsterisk{}_{((\ell, i), G) \in m} G * \exists v.~ (G \guards[\maskHeap] (\ell \pointsto v))
\end{align*}}%
Here, $\countsAgree(\sigma,m)$ means that for each $\ell$, the number of entries of the form $(\ell, \_)$
in $m$ is equal to the read-count for $\ell$ in $\sigma$.

The point of all this is that, when we are in the middle of a non-atomic read with a guard
proposition $G$, we can save $G$ in $\HeapInterp(\sigma)$ during the intermediate step.

Now, observe that:
\begin{itemize}
  \item When we have a shared $\ell \pointsto v$, we can deduce that $\sigma(\ell)$
      is not in the $\stateWriting$ state.
  \item When we have exclusive access to $\ell \pointsto v$, we can further deduce
      that $\sigma(\ell)$ is in the $\stateReading{0}$ state.
      To do this, assume by contradiction we are in the $\stateReading{n}$ state for $n > 0$.
      Then, there is at least one $\ell$-entry in $m$, so we can obtain a guard proposition
      $G$ where $G \guards (\ell \pointsto v)$. This is a contradiction because we have
      exclusive ownership of $\ell \pointsto v$.
\end{itemize}
This allows us to prove that we do not reach a stuck state when performing the operations.

For the proofs:
\begin{itemize}
    \item To prove \ruleref{Heap-Write-NA}, we perform the first execution step,
        moving from $\stateReading{0}$ into $\stateWriting$, and exchanging
        $\ell \pointsto v$ for $\pointstoInter{\ell}{v}$.
        In the second execution step, we switch back and re-obtain $\ell \pointsto v$.
    \item To prove \ruleref{Heap-Read-Shared-NA}, recall first that it is equivalent to,
\[
  \{ G * (G \guards[\maskHeap] (\ell \pointsto v)) \} ~ \langbangNA\ell ~ \{ v'.~ (v = v') * G \}_{\maskHeap}
\]
        To show this, we first pick some arbitrary fresh index $i$
        and add $[(\ell, i) \mapsto G]$ to the map $m$.
        Since we are moving from $\stateReading{n}$ to $\stateReading{n+1}$,
        this means the counts still agree.
        Furthermore, it means we exchange $G$ for $\ownGhost{\gammaAux}{\authfrag [ (\ell, i) \mapsto \latertinj(G) ]}$.
        This lets us ``remember'' what $G$ is, so we can re-obtain $G$ when we perform
        the second step.

\end{itemize}
Of course, \ruleref{Heap-Write-SC} and \ruleref{Heap-Read-Shared-SC} are simpler to prove,
since the corresponding heap operations are atomic.

\end{document}